\documentclass[useAMS,usenatbib,usegraphicx]{mn2e}%,referee
\usepackage{url}
\usepackage{graphicx}
\usepackage{enumerate}
\usepackage{amssymb,amsmath}
\usepackage{color}
\title[Far IR LF of UV-selected galaxies]{HerMES: Unveiling obscured
  star formation -- the far infrared luminosity function of
  ultraviolet-selected galaxies at
  $\boldsymbol{z\sim1.5}$}
\author[S.~Heinis et al.]
{\parbox{\textwidth}{\raggedright S.~Heinis,$^{1}$\thanks{E-mail: \texttt{sebastien.heinis@oamp.fr}}
V.~Buat,$^{1}$
M.~B{\'e}thermin,$^{2,3}$
H.~Aussel,$^{2}$
J.~Bock,$^{4,5}$
A.~Boselli,$^{1}$
D.~Burgarella,$^{1}$
A.~Conley,$^{6}$
A.~Cooray,$^{7,4}$
D.~Farrah,$^{8}$
E.~Ibar,$^{9}$
O.~Ilbert,$^{1}$
R.J.~Ivison,$^{9,10}$
G.~Magdis,$^{11}$
G.~Marsden,$^{12}$
S.J.~Oliver,$^{13}$
M.J.~Page,$^{14}$
G.~Rodighiero,$^{15}$
Y.~Roehlly,$^{1}$
B.~Schulz,$^{4,16}$
Douglas~Scott,$^{12}$
A.J.~Smith,$^{13}$
M.~Viero,$^{4}$
L.~Wang$^{13}$ and
M.~Zemcov$^{4,5}$}\vspace{0.4cm}\\
\parbox{\textwidth}{\raggedright $^{1}$Aix Marseille Universit\'e, CNRS, LAM (Laboratoire d'Astrophysique de Marseille) UMR 7326, 13388, Marseille, France\\
$^{2}$Laboratoire AIM-Paris-Saclay, CEA/DSM/Irfu - CNRS - Universit\'e Paris Diderot, CE-Saclay, pt courrier 131, F-91191 Gif-sur-Yvette, France\\
$^{3}$Institut d'Astrophysique Spatiale (IAS), b\^atiment 121, Universit\'e Paris-Sud 11 and CNRS (UMR 8617), 91405 Orsay, France\\
$^{4}$California Institute of Technology, 1200 E. California Blvd., Pasadena, CA 91125, USA\\
$^{5}$Jet Propulsion Laboratory, 4800 Oak Grove Drive, Pasadena, CA 91109, USA\\
$^{6}$Center for Astrophysics and Space Astronomy 389-UCB, University of Colorado, Boulder, CO 80309, USA\\
$^{7}$Dept. of Physics \& Astronomy, University of California, Irvine, CA 92697, USA\\
$^{8}$Department of Physics, Virginia Tech, 910 Drillfield Drive, Blacksburg, VA 24061, USA\\
$^{9}$UK Astronomy Technology Centre, Royal Observatory, Blackford Hill, Edinburgh EH9 3HJ, UK\\
$^{10}$Institute for Astronomy, University of Edinburgh, Royal Observatory, Blackford Hill, Edinburgh EH9 3HJ, UK\\
$^{11}$Department of Astrophysics, Denys Wilkinson Building, University of Oxford, Keble Road, Oxford OX1 3RH, UK\\
$^{12}$Department of Physics \& Astronomy, University of British Columbia, 6224 Agricultural Road, Vancouver, BC V6T~1Z1, Canada\\
$^{13}$Astronomy Centre, Dept. of Physics \& Astronomy, University of Sussex, Brighton BN1 9QH, UK\\
$^{14}$Mullard Space Science Laboratory, University College London, Holmbury St. Mary, Dorking, Surrey RH5 6NT, UK\\
$^{15}$Dipartimento di Astronomia, Universit\`{a} di Padova, vicolo Osservatorio, 3, 35122 Padova, Italy\\
$^{16}$Infrared Processing and Analysis Center, MS 100-22, California Institute of Technology, JPL, Pasadena, CA 91125, USA}}

\begin{document}

\date{}

\maketitle
\begin{abstract}
We study the far-infrared (IR) and sub-millimeter properties of a
sample of ultraviolet (UV) selected galaxies at $z \sim 1.5$. Using
stacking at 250, 350 and 500\,$\mu$m from \textit{Herschel} Space
Observatory SPIRE imaging of the COSMOS field obtained within the
HerMES key program, we derive the mean IR luminosity as a function of
both UV luminosity and slope of the UV continuum $\beta$. The IR to UV
luminosity ratio is roughly constant over most of the UV luminosity
range we explore. We also find that the IR to UV luminosity ratio is
correlated with $\beta$. We observe a correlation that underestimates
the correlation derived from low-redshift starburst galaxies, but is
in good agreement with the correlation derived from local normal
star-forming galaxies. Using these results we reconstruct the IR
luminosity function of our UV-selected sample. This luminosity
function recovers the IR luminosity functions measured from IR
selected samples at the faintest luminosities ($L_{\rm{IR}} \sim
10^{11}\,{\rm L}_{\sun}$), but might underestimate
them at the bright-end ($L_{\rm{IR}} \ga 5\times10^{11}\,{\rm
  L}_{\sun}$). For galaxies with $10^{11}<L_{\rm{IR}}/{\rm
  L}_{\sun}<10^{13}$, the IR luminosity function of a UV selection
recovers (given the differences in IR-based estimates) 52-65 to 89-112 per cent of the
star-formation rate density derived from an IR selection. The cosmic
star-formation rate density derived from this IR luminosity function
is 61-76 to 100-133 per cent of
the density derived from IR selections at the same
epoch. Assuming the latest \textit{Herschel} results
  and conservative stacking measurements, we use a toy model to fully
  reproduce the far IR luminosity function from our UV selection at
  $z\sim 1.5$. This suggests that a sample around 4 magnitudes deeper
  (i.e. reaching $u^* \sim 30$\, mag) and a large
  dispersion of the IR to UV luminosity ratio are required.

\end{abstract}

%Using a toy model, we propose that fully
%reproducing the far IR luminosity function from our UV selection at
%$z\sim 1.5$ requires a sample around 4 magnitudes deeper
%(i.e. reaching $u^* \sim 30$\, \textcolor{blue}{mag}) and a large
%dispersion of the IR to UV luminosity ratio.


\section{Introduction}
%Impact of dust extinction on constraints of evolution of Cosmic star
%formation evolution; how to derive accurate SFRs and avoiding
%selection biases ?  Refers to: Buat at al., Martin et al. 2005-2007;
%Lagache et al. 1999, Bothwell et al. 2011, Reddy et al. ..

Star-formation is one of the main properties used to trace galaxy
formation and evolution. Our ability to constrain the mechanisms that
drive galaxy evolution hence depends to a large extent on our ability
to measure accurate star-formation rates for samples of galaxies. A
number of star-formation tracers are routinely used, from the
strengths of spectral lines to broad-band measurements
\citep{Kennicutt_1998}. Broad-band measurements offer the advantage
that estimates can be derived for a large number of galaxies with
minimal spectroscopic follow-up, albeit at the expense of possible
contamination by strong spectral features
\citep[e.g.][]{Smail_2011}. One of the main star-formation rate
tracers of this kind comes from the ultraviolet (UV) range of the
spectrum, where most of the energy is emitted by young stars
\citep[ages $\sim 10^7 {-} 10^8$\,Myr; see e.g.][]{Martin_2005a}. The
UV has been widely used over a large redshift range to infer the
cosmic star-formation density, from $z=7$ to $z=0$
\citep[e.g.][]{Schiminovich_2005, Bouwens_2009}.

Optical/near infrared (IR) observations probe the rest-frame UV for
samples of galaxies at high redshift, hence it has been the primary
choice for constraining the star-formation activity of the Universe at
early epochs from large samples. However, interstellar dust, which is
a byproduct of star-formation, makes the measurement of star-formation
activity challenging at these wavelengths. Dust grains scatter or
absorb the light emitted by young stars; hence only a fraction of the
energy output from star-formation is observable in the UV.  Dust
grains re-emit this energy over the full IR range $8-1000\,\mu$m. One
way to estimate the dust attenuation in the UV and to assess the
selection bias inherent to the UV is then to study the far IR
properties of UV-selected galaxies.

% and to devise ways of correcting for it
A number of studies have used this approach to characterise the amount
of dust attenuation. In the local Universe, the star-formation rate
density is roughly equally divided between UV and IR contributions
\citep{Martin_2005b, Bothwell_2011}. At earlier epochs, however, the
fraction of star-formation rate density, which is directly measurable
using the UV continuum, decreases from 44 per cent ($z=0$) to roughly
15 per cent at $z\sim 1$ \citep{Takeuchi_2005, Tresse_2007}, while it
might increase slightly to 20 per cent at $2<z<3.5$
\citep{Reddy_2008}, and even to higher values in the early Universe
\citep{Bouwens_2010}.

%fraction sfr_uv/(sfr_uv+sfr_ir)
%Reddy et al. 2008
%1.9<z<2.7: 19.4%
%2.7<z<3.4: 22%
To overcome this drawback and the lack of deep IR data, it is common
to use empirical recipes to correct UV for dust attenuation. The most
well-known is the relation between the slope of the UV continuum and
the ratio of the luminosities in the IR and the UV
\citep{Meurer_1999}. The slope of the UV continuum can be derived from
rest-frame UV colours, and hence is convenient for estimating star
formation rates at high redshifts when a limited wavelength range is
available \citep{Schiminovich_2005, Bouwens_2009}.

However, this recipe does encounter several pitfalls: it has been
derived from local starburst galaxies, and it might not be valid for
more normal star-forming galaxies \citep{Boissier_2007, Cortese_2006,
  Munoz-Mateos_2009, Hao_2011}. Moreover, the relation between dust
attenuation and UV slope depends on the selection criteria
\citep{Buat_2005,Seibert_2005}, and might also be sensitive to
star-formation history \citep{Kong_2004,Panuzzo_2007}, as well as dust
properties \citep{Inoue_2006} and dust geometry \citep{Calzetti_2001}.

It is hence of particular importance to follow the evolution with
redshift of the IR properties of UV-selected galaxies, in order to
characterize the biases inherent in such selection, and also to
examine the validity of the empirical recipes commonly used to correct
for dust attenuation.

In this context a new era started with the availability of data from
the \textit{Herschel}\footnote{\textit{Herschel} is an ESA space
  observatory with science instruments provided by European-led
  Principal Investigator consortia and with important participation
  from NASA.} telescope \citep{Pilbratt_2010}. Indeed, while
\textit{Spitzer} data uncovered the dusty star-formation history of
the Universe up to $z\simeq 1$ \citep{Lefloch_2005}, at higher
redshifts large extrapolations are needed to estimate IR luminosities
from 24$\,\mu$m data, which could lead to systematic errors
\citep{Bavouzet_2008a}, as they do not probe the peak of the dust
emission. Based on \textit{Herschel} data, \citet{Elbaz_2010} showed,
for instance, that using mid-IR data at $z>1.5$ leads to an
overestimation of the total IR luminosity.  Another important feature of
the submillimeter wavelength range is that the contribution of Active
Galactic Nuclei to the galaxy Spectral Energy Distribution (SED) is
generally outweighed by the star-formation component for $\lambda \ga$
30\,$\mu$m \citep{Hatziminaoglou_2010}.

%do it at higher redshifts \ldots with unprecedented statistics \ldots
%with more accurate measurements of $L_{IR}$ see Elbaz et al; also no
%contamination from AGN in the wavelength range probed by herschel
In this paper, we focus on a UV-selected sample at $z\sim 1.5$ to
study with unprecedented statistics the far IR properties of
UV-selected galaxies using \textit{Herschel} data for more accurate
measurements of IR luminosities. Given the confusion-limited nature of
these data, we rely on a stacking analysis to derive the mean IR
luminosities for different classes of object
\citep[e.g.][]{Bethermin_2012, Hilton_2012,Viero_2012b}.

The paper is organized as follows. We start by presenting the sample
construction and the need for stacking (Section. \ref{sec_data}). In
Section \ref{sec_stacking} we present our methods for stacking
measurements and corrections for biases. Section \ref{sec_stack_Luv}
shows our stacking results as a function of UV luminosity and Section
\ref{sec_stack_beta} as a function of the slope of the UV
continuum. In Section \ref{sec_LF} we reconstruct the total IR
luminosity function of our UV-selected sample using the stacking
results, and examine the implications of these results for the
measurement of the cosmic star-formation density from UV and IR
selected samples. We discuss these results in Section
\ref{sec_discussion} and present our conclusions in Section
\ref{sec_conclusion}.

In this paper, we use a standard cosmology with $\Omega_{\rm m} =
0.3$, $\Omega_{\Lambda} = 0.7$ and $H_{0} = 70$\,km
s$^{-1}$Mpc$^{-1}$, denote FUV and IR luminosities as $\nu L_{\nu}$, and use AB magnitudes.

%\citet{Capak_2007} built multi-bands catalogs by matching the
%PSFs to the band with worst quality and then detecting objects in the
%$i$-band image. While such procedure is appropriate for multi-band
%data, it is not optimal for our study which is based only on
%$u^{\ast}$ and $V_J$ bands. Hence we did not use \citet{Capak_2007}
%catalogs but 
\section{Data sample}\label{sec_data}
We use optical imaging of the COSMOS field from \citet{Capak_2007} in
the $u^{\ast}$-band (obtained at CFHT; depth:
26.4\,mag at 5$\sigma$ for a 3\,arcsec aperture) and
$V_J$-band (from Subaru; depth: 26.6\,mag, also at
5$\sigma$ for a 3\,arcsec aperture). We generated catalogues from the
$u^{\ast}$ and $V_J$ images using SExtractor \citep{Bertin_1996} in
order to select galaxies directly in the $u^*$-band, and obtain
accurate total fluxes. Comparison with the photometry from
\citet{Capak_2007} shows good agreement. Hereafter all quoted
magnitudes are corrected for Galactic extinction using dust maps from
\citet{Schlegel_1998}.

%We measured the completeness of this catalogue by injecting 1000
%point sources \citep[FWHM = $0.9$\,arcsec,][]{Capak_2007} into the
%$u^{\ast}$ image with random positions, and a magnitude distribution
%similar to the observed counts, in the range $21<u^{\ast}<27$. We
%performed source extraction with SExtractor on this new image using
%the same parameters as the ones used to generate the $u^{\ast}$-band
%catalogue, and kept track of the input sources that were
%recovered. We performed aperture photometry on the image with sources
%added, (using an aperture diameter of 2\,arcsec, twice the FWHM in
%the $u^{\ast}$ band), and then cross-matched the catalogue with the
%input one using a 0.5\,arcsec search radius. We performed this
%process 50 times to have sufficient statistics to
%quantify the effects of incompleteness on stacking (see
%Sect. \ref{incomp}). We estimate that the completeness is $83\pm2$ per
%cent at the limiting magnitude $u^{\ast}=26$
%(Fig. \ref{fig_completeness}).

%Fathi et al. use ACS F850LP filter (effective lambda is 9194 A)
We estimated the incompleteness by injecting 1000 fake sources in the
$u^{\ast}$ image with random positions.  We injected sources only
within areas not masked for edges or bright stars. We assumed that the
objects in our UV-selected sample are pure exponential disks. We
assumed that the disk scale length for galaxies with $L_{\textrm{FUV}}
= L^* \sim 10^{10}\rm{L}_{\odot}$ is equal to 3 kpc. This value is 50
per cent larger than the one from \citet{Fathi_2012}, determined at
$0.3<z<2.0$ in the $z$-band (restframe wavelength 4600\,\AA~at $z=1$),
in order to account for the fact that disks are larger at UV
wavelengths. The value we assumed is in agreement with predictions
from \citet{Boissier_2001}, as well as with the UV restframe
measurements from \citet{Ferguson_2004}, taking into account that the
mean luminosity of their sample is around $5\times L^*_{\textrm{FUV}}
(z=1.5)$. We further assumed that the disk scale length varies with
luminosity as $L^{1/3}$, as observed in the local Universe
\citep{deJong_2000}. We generated fake objects following the observed
joint distributions of magnitude, redshift, and minor to major axis
ratio; we also allowed random position angles. Then for each fake
object we could infer its UV luminosity and its disk scale length. We
injected these objects in the $u^{\ast}$ image after convolving them
with the $u^{\ast}$-band PSF \citep[FWHM =
  $0.9$\,arcsec,][]{Capak_2007}. We performed source extraction with
SExtractor on this new image using the same parameters as the ones
used to generate the $u^{\ast}$-band catalogue, but using this time
the \texttt{ASSOC} mode to cross-match directly the detections with
our input fake sources list, with a 0.5\,arcsec search radius. We
performed this process 1500 times to have sufficient
statistics to quantify the effects of incompleteness on stacking (see
Sect. \ref{incomp}). In order to test the impact of
  real sources on the flux estimation of the fake sources, we
  cross-matched the fake sources we recovered with the full $u^*$-band
  catalog. We then rejected fake sources whose magnitudes are
  perturbed by closeby real sources, based on the angular separation
  and the difference between the input and the recovered magnitude.

\begin{figure}
  \includegraphics[width=\hsize]{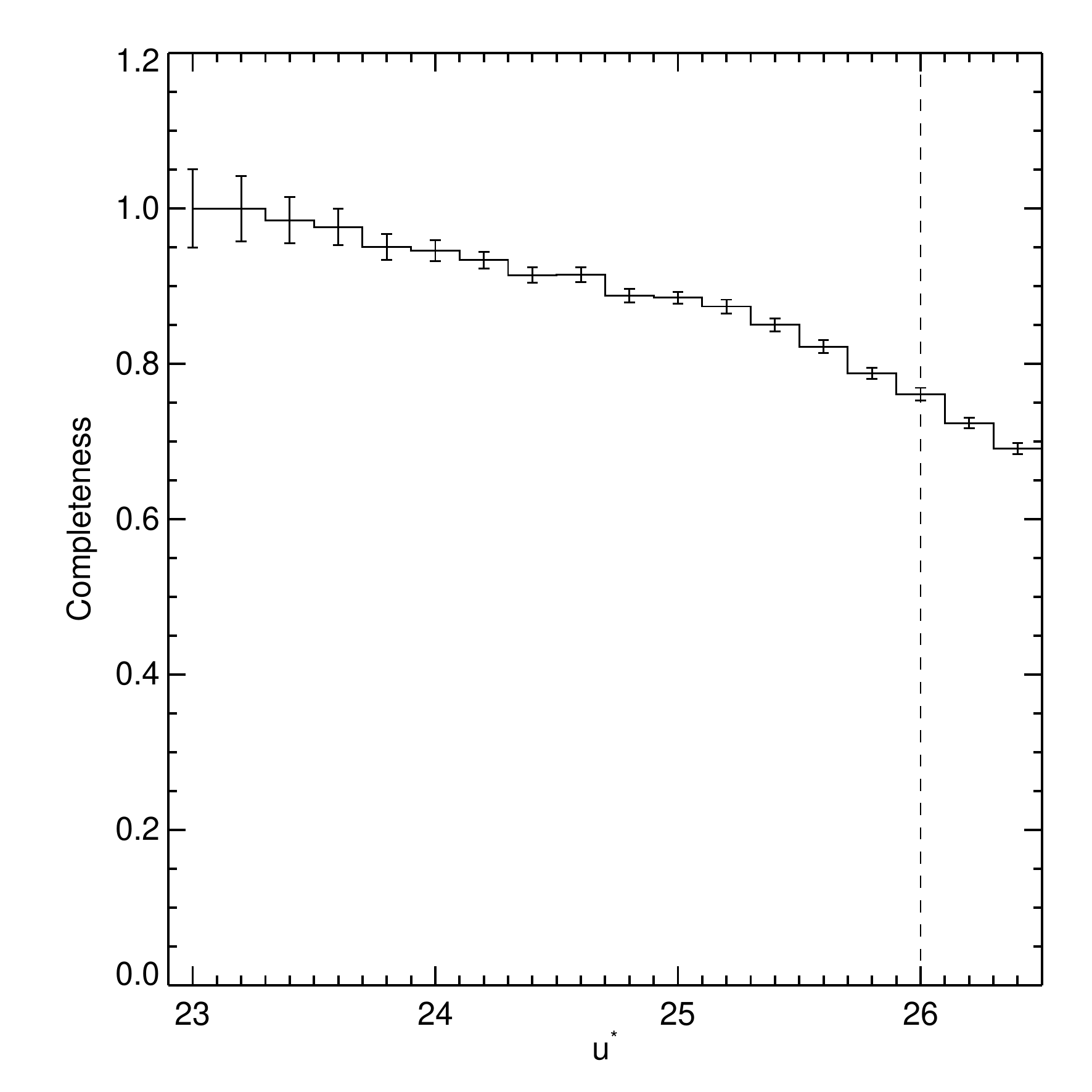}
\caption{Completeness in the $u^{*}$ band for the UV-selected
  catalogue. Error bars are the errors on the mean. The vertical
  dashed line shows the magnitude limit we adopt.} %The completeness
%at $u^{*}=23.6$ is larger than 1 because of a few sources which have
%been extracted brighter than their input magnitudes.
\label{fig_completeness}
\end{figure}

We estimate that the completeness is $76\pm1$ per
cent at the limiting magnitude $u^{\ast}=26$\,mag
(Fig. \ref{fig_completeness}).

%\footnote{The completeness limit we
%  derive is in agreement with that available from the CFHTLS-T0006
%  release document, which quotes a 80 per cent completeness at
%  $u^*=25.82\pm 0.1$, for the Deep field D2, with half the exposure
%  time we use here.}

%The accuracy at $1\sigma$ in $(1+z)$ is $\sim 0.06$ at $i_{\rm AB}^{+}
%\sim24$ up to $z\sim 2$.

We cross-matched our sample with an updated version of the photometric
redshift catalogue of \citet[][v.2.0]{Ilbert_2009}. This version
differs from the original \citet{Ilbert_2009} catalogue by the
inclusion of additional near-IR photometry
\citep[YJHK,][]{McCracken_2012} which improves the accuracy of the
photometric redshifts, in particular in the redshift range we are
interested in. 98.5 per cent of the sources from our sample have a
counterpart in the \citet{Ilbert_2009} catalogue within a 1\,arcsec
search radius.

We built a rest-frame UV-selected catalogue considering objects with
$1.2<z_{\rm phot}<1.7$ and $u^{\ast}<26$\,mag. The
average photometric redshift accuracy at $1\sigma$ in $(1+z)$ is 0.04
for this sample, and the mean redshift is $\langle z_{\rm phot}\rangle
= 1.43$. At this redshift, the effective wavelength of the $u^*$-band
filter corresponds to 1609\,\AA, which is in the far-UV (FUV)
rest-frame. Our final catalogue contains 42,184 objects over
1.68\,deg$^2$, after masking of edges and areas around bright stars.

%\begin{figure}
%  \includegraphics[width=\hsize]{./plots/nz.eps}
%\caption{Redshift distribution of the UV-selected catalog.}
%\label{fig_nz}
%\end{figure}

We use \textit{Herschel}-SPIRE \citep{Griffin_2010, Swinyard_2010}
imaging at 250 (FWHM $= 18.15$\,arcsec), 350 (FWHM $= 25.15$\,arcsec),
and 500\,$\mu$m (FWHM $= 36.3$\,arcsec) of the COSMOS field obtained
as part of the \textit{Herschel} Multi-Tiered Extragalactic
Survey\footnote{\url{http://hermes.sussex.ac.uk}}
\citep[HerMES,][]{Oliver_2012} programme. We use here the images
produced by the SMAP pipeline \citep{Levenson_2010, Viero_2012a}. The
effective area (after removing masked regions) of the overlap between
$u^{\ast}$-band and \textit{Herschel}-SPIRE images is 1.5\,deg$^2$;
38,074 galaxies from our UV-selected catalogue are within this area.

% Note that we expect to find by chance $\sim 2.4$ UV sources within
% such a radius.

We performed cross-matching between our UV-selected catalogue and the
SCAT \citep{Smith_2012} blind detections at 250, 350 and 500\,$\mu$m,
using a 5\,arcsec search radius. Considering as `detected' only
\textit{Herschel}-SPIRE sources with fluxes larger than the confusion
limit \citep[at 5$\sigma$, 24.0\,mJy at 250\,$\mu$m, 27.5\,mJy at 350\,$\mu$m,
  and 30.5\,mJy at 500\,$\mu$m,][]{Nguyen_2010}, we find that less than 1
per cent of the UV sources are detected at 250, 350, and 500\,$\mu$m.
This result implies that we need to use a stacking analysis in order
to study in a statistical way the IR properties of the UV-selected
galaxies in our sample. In the following, we include all UV-selected
galaxies within the HerMES footprint in the stacking analysis, whether
they are detected at the \textit{Herschel}-SPIRE wavelengths or not.
Excluding UV-selected galaxies detected at SPIRE wavelengths from the
stacking input lists does not significantly impact our results.

We further use this sample of UV-selected galaxies detected at SPIRE
wavelengths to determine the dispersion in $L_{\rm IR}/L_{\rm FUV}$ as
a function of $L_{\rm FUV}$. This sample has a mean infrared
luminosity $\langle L_{\rm IR} \rangle =10^{12}\rm{L}_{\odot}$, is
slightly brighter in UV than our full sample ($\langle L_{\rm FUV}
\rangle =2.6\times 10^{10}\rm{L}_{\odot}$, compared to
$10^{10}\rm{L}_{\odot}$ for the full sample), and has a mean IR to UV
luminosity ratio of $\langle L_{\rm IR}/L_{\rm FUV}\rangle = 66$.

\section{Stacking measurements}\label{sec_stacking}
%We present in this section the methods we use to measure fluxes via
%stacking. 
We use the IAS library \citep{Bavouzet_2008b,
  Bethermin_2010a}\footnote{\url{http://www.ias.u-psud.fr/irgalaxies/files/ias_stacking_lib.tgz}}
to perform the stacking. We use the calibrated 250, 350 and
500\,$\mu$m images, and we do not attempt to clean the image of any
detected sources at SPIRE wavelengths. 

%Note that doing so would bias the measurements, and induce an
%underestimation of the stacked flux \citep{Marsden_2009}; this holds
%under the assumption that the stacking is performed on an un-clustered
%population.
%This underestimation would be larger for a lower flux cut for cleaning
%detected sources.

% Computing the stacking from the median instead of the mean yields
% similar results.

For a given stacking measurement, we generate both a postage stamp
image and a radial profile, using the mean of the individual images
that are included in the stacking. We derive errors on
the profiles by bootstrap resampling.

The background for the stack images is usually considered constant and
determined by the average value of the stack profile at large
distances from the centre of the stack image. However, a number of
effects can yield a non homogeneous background; we correct our
stacking measures for two effects which have an impact on the
background: the incompleteness of the input catalogue, and the
clustering of the input galaxies. These corrections are based on the
work of \citet{Bavouzet_2008b} and \citet{Bethermin_2010b}.

\subsection{Correcting for stacking bias}\label{incomp}

\begin{figure*}
  \includegraphics[width=\hsize]{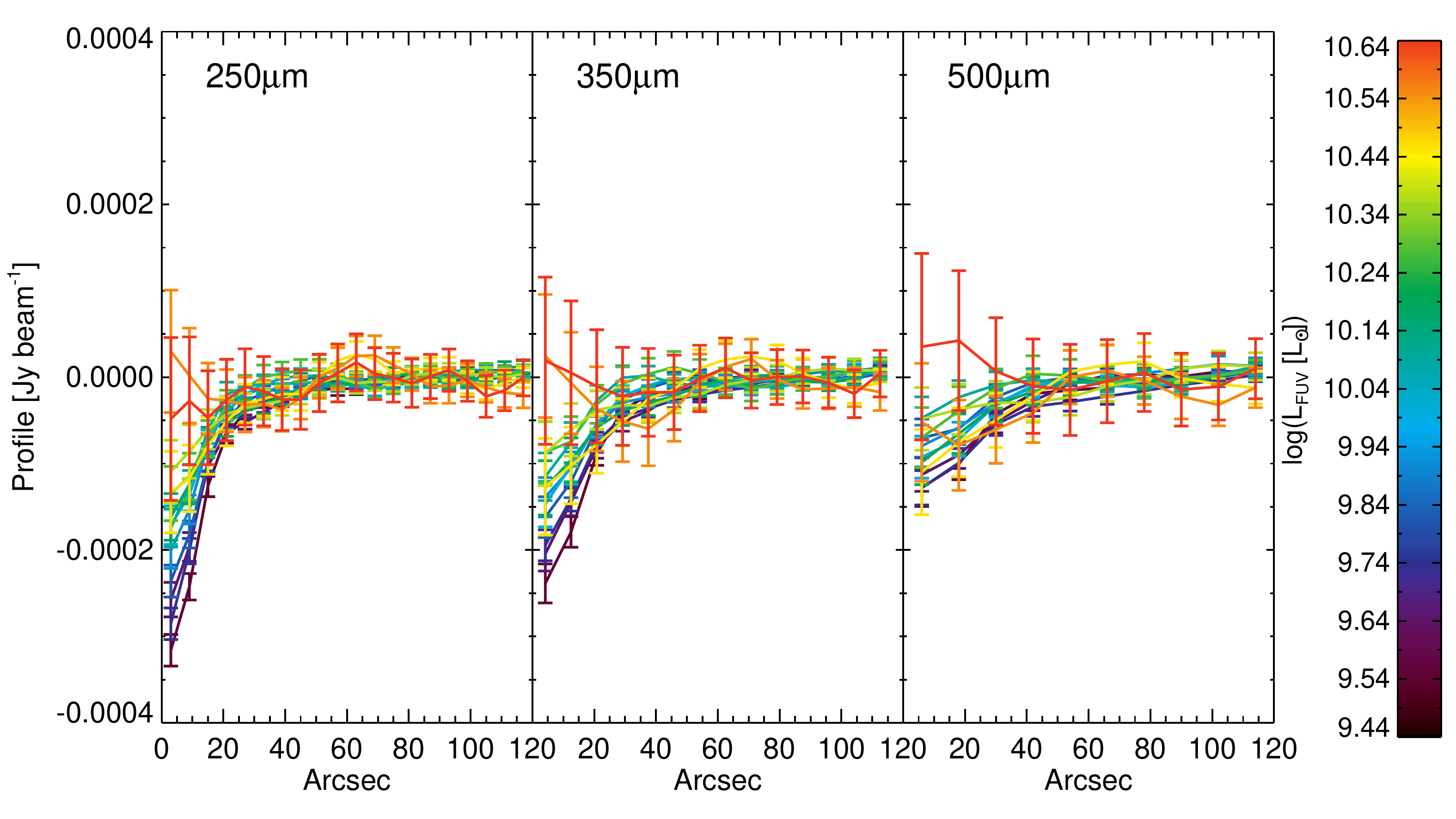}
\caption{Profiles of the stacking at 250, 350 and 500\,$\mu$m of
  recovered fake sources in bins of $L_{\rm FUV}$ luminosity.}
\label{fig_stack_incompleteness}
\end{figure*}

While limited, the incompleteness at the faint-end of the input
catalogue can have an impact on the stacking measurements, for
instance if it is partly related to the local geometry during the
detection process. In particular, the detection efficiency for faint
objects is lower in dense areas of the image. As we will see, this
effect can also be described by a clustering term under the form of a
cross correlation between sources of different $u^*$ band fluxes. 

If we stack a population of objects which is randomly distributed in
the sky, we will get a flat background. On the other hand, if there is
a bias introduced by the detection process, we will miss the
contribution of the objects which are not recovered by the source
extraction. The basis of this correction is to measure the actual
background of the stacking for a given class of input sources. 
%To
%test whether any effect of this kind is present in our case, we used
%the results from the incompleteness study described above
%(Sect. \ref{sec_data}).  

%As expected, at large scales ($\theta > 50$ arcsec), the profiles are relatively consistent with zero.
% averaged over 100 iterations 
% are the errors on the mean from the individual profiles of each iteration.

We assume here that the stacking bias effects are related to the UV
luminosity of the objects.  In other words, we consider the actual
$L_{\rm FUV}$ distribution for each class of galaxy to correct for
this bias. To quantify the impact of the stacking bias, we use the
fake sources created to estimate the completeness of our catalogue
(see Sect. \ref{sec_data}). We stack the fake sources recovered by the
detection process with the same UV luminosity distribution as the
class of galaxy we are considering. We show in
Fig. \ref{fig_stack_incompleteness} the radial profiles of the
stacking as a function of $L_{\rm FUV}$. These profiles are based on a
sample of fake sources with roughly 25 times the actual number of
galaxies in our sample. The error bars on these profiles are obtained
through bootstrap resampling. For faint objects ($L_{\rm
  FUV}<10^{10}\,{\rm L}_{\sun}$), at smaller scales, the profiles are
actually lower than zero. Note that the amplitude of this effect
increases for fainter objects.

This result is related to the well-known effect that the detection
efficiency is lower in dense areas, and in particular for faint
objects which are close to brighter ones. If all fake sources were to
be recovered by the UV source extraction, the profile of their
stacking at the \textit{Herschel}-SPIRE wavelength would be zero, as
their input distribution is independent of that of the real
sources. However, the faint sources which are close to bright real
sources in the $u^*$ image are not completely recovered. This effect
is more important for fainter sources, as well as for smaller
distances with respect to bright sources. Hence, the stacking of faint
objects is missing the contribution to the background of the ones
which happen to be closer to UV-bright sources, and then the
background at small scales is lower. 

%In some cases, the radial profiles we measure are positive at small
%scales; the profile for the brightest bin is negative at small scales
%which is also at odds with what is expected. We believe this is caused
%by a few artefacts created during the detection process of the fake
%sources; note however (see Fig. \ref{fig_fluxes_cor}) that this has a
%completely negligible impact on our results.

%This effect is related to the fact that the detection of faint objects
%in the UV data is less efficient around brighter ones. In practice,
%the probability of finding a faint object around a bright one
%decreases for angular distances lower than about 20\arcsec
%\citep{Bavouzet_2008b}. In turn, this implies that the contribution to
%the stack of faint objects located around bright ones is missed, and
%hence the actual background lower. This causes an
%\textit{underestimation} of the average flux.

In order to correct our stacking measurement for this effect, we
subtract these profiles, without smoothing, from the profiles of the
stacked images. We add in quadrature the errors of these profiles to
the errors of the stacked image profiles.

%For each stacking measure, we generate the incompleteness profile by
%stacking recovered fake sources with the same $u^{*}$ magnitude
%distribution than the magnitude distribution of the actual galaxies
%involved in the stack measure.

%The stacked image enables to measure directly the background at large
%scales; while this value of the background is correctly recovered, the
%actual value of the background at small scales is not.

%\textbf{effect at faint mags:} background is not uniform; an uniform
%bacgkroung is valid at large scales around the stacked sources, but
%the actual background is actually lower at small distances around the
%stacked sources because the detection of faint sources is less
%efficient around bright sources, while when faint sources located
%further to bright ones are better recovered, and hence these bright
%sources do contribute to the background.

We show as squares in Fig. \ref{fig_fluxes_cor} the ratio of the flux
densities corrected from the stacking bias to the flux densities
measured with PSF-fitting, assuming a constant background, as a
function of $L_{\rm FUV}$. The amplitude of the correction is maximal
at faint luminosities (around 2 times the flux from direct PSF-fitting
at $L_{\rm FUV}\sim 6\times10^{9}\,{\rm L}_{\sun}$) then decreases
with luminosity, to be negligible for $L_{\rm FUV}\ga
2\times10^{10}\,{\rm L}_{\sun}$.

%Compared to a PSF-fitting flux measurement with a constant background, this correction yields an increase of the flux of at most 35\% at faint magnitudes ($u<24$),  and is negligible for magnitudes brighter than $u>24.$.

%so there is a non zero probability that there are other galaxies
%belonging to the same class within the stacking field of view

\subsection{Correcting for clustering of the input catalogue}\label{sec_clustering}
We use the formalism developped by \citet{Bavouzet_2008b} and
\citet{Bethermin_2010b} in order to take into account the impact on
the measured flux from the clustering of the population under
study. If the input population is uniformly distributed on the sky,
the measured stacking is just the average flux of the population at
the stacked wavelength. In practice, galaxies are clustered, so there
is an excess of probability to find another galaxy of the sample
within the beam, compared to the value derived from a randomly
distributed population. This yields an \textit{overestimation} of the
flux. The probability is proportional to the angular correlation
function of the class of galaxies under study. The two dimensional
profile of the resulting stacking can then be written as:

\begin{equation}\label{eq_stack_wtheta}
  I(\theta, \phi) = \overline{S}\times \textrm{PSF}(\theta, \phi) + a\times \big[w(\theta, \phi)\ast
  \textrm{PSF}(\theta, \phi)\big]\textrm{.}
\end{equation}
Here $\overline{S}$ is the average flux, PSF$(\theta, \phi)$ is the
Point Spread Function at the stacked wavelength, $w(\theta, \phi)$ is
the angular autocorrelation function of the input population; the
symbol $\ast$ denotes a convolution, and $a$ is a parameter relating
to the density of the input population and their contribution to the
cosmic background. The effect of the clustering is to
  add to the profile a component which is broader than the pure PSF
  \citep[see e.g.][their fig. 3]{Bethermin_2012}.

To correct our stacking measures for clustering, we adjust the radial
profile of each stacked image over 120\,arcsec following
eq.~\ref{eq_stack_wtheta}, with $\overline{S}$ and $a$ as free
parameters, and using the autocorrelation function $w(\theta)$ of the
sample. We obtain the best value of $\overline{S}$ by marginalising
the two dimensional probability of $(\overline{S}, a)$ over $a$, using
$\chi^2$ statistics.

In this paper we perform stacking in bins of UV luminosity and the
slope of the UV continuum ($\beta$). We measure $w(\theta)$ using the
method of \citet{Szapudi_2005}, and fit it with a power law $w(\theta)
= A_{w}\theta^{-\delta}$, correcting for the integral constraint
following \citet{Roche_1999}. We find that the correlation function is
well modelled with $\delta = 0.64\pm 0.07$. Note that we consider $a$
as a free parameter in eq. \ref{eq_stack_wtheta}, so our results do
not depend on the actual value of $A_{w}$.

%For the full UV-selected sample we find $A_w = 0.003\pm 0.001$ and
%$\delta = 0.63\pm 0.06$. For a subsample with $1.2<z<1.3$, used to
%stack as function of the $\beta$ slope, we measure $A_w =
%0.009^{+0.004}_{-0.003}$ and $\delta = 0.64\pm 0.07$. 

The angular correlation function depends on UV luminosity \citep[see
  e.g.][]{Giavalisco_2001, Heinis_2007, Savoy_2011}, while there is no
evidence that it depends on $\beta$. We checked that while there are
some variations in amplitude and slope of the correlation function
with UV luminosity, they are not sufficient in this context to
significantly change the correction for the flux determination. We
consider only the best fit to the correlation function from the full
sample hereafter.

%When
%stacking as a function of $L_{\rm FUV}$, we consider only the best fit
%to the correlation function from the full sample, and when stacking as
%a function of $\beta$ the best fit to the correlation function of the
%corresponding subsample.

%Indeed the signal to noise ratio of the $w(\theta)$ measures
%per UV luminosity bins or $\beta$ bins is significantly lower than the
%one of the full sample. Using such measures would in turn increase
%errors on the fluxes obtained from stacking. 

\begin{figure}
  \includegraphics[width=\hsize]{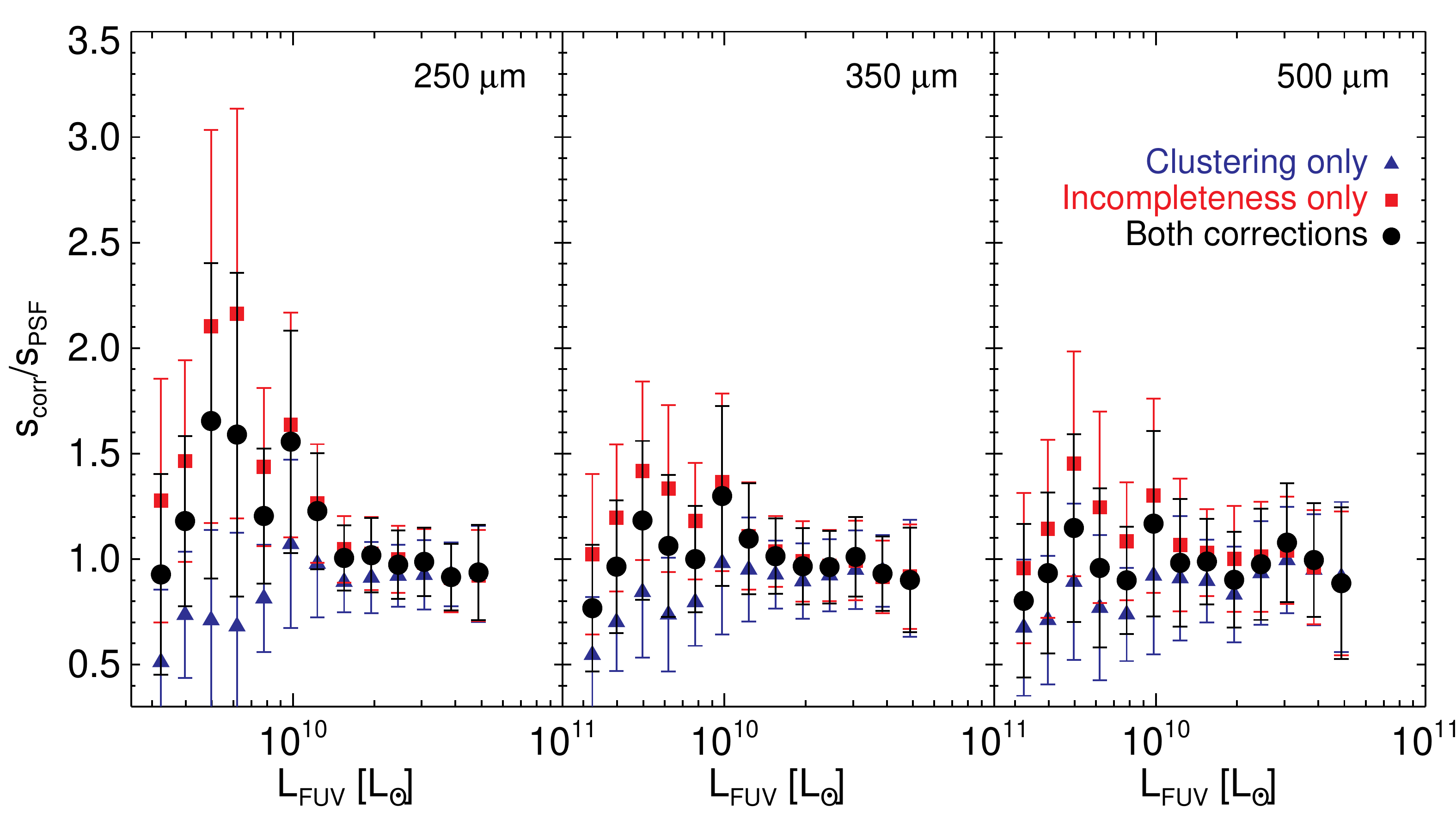}
\caption{Amplitude of the correction applied to stacked flux densities
  as a function of $L_{\rm FUV}$, at 250, 350 and 500\,$\mu$m. We show
  here the ratio of the corrected stacked flux density to the flux
  density measured by a simple PSF-fitting. The triangles show this
  ratio for the correction due to the clustering of the input
  catalogue, the squares are the correction applied because of the
  incompleteness of the input catalogue, and the circles show the
  ratio for the total correction including clustering and
  incompleteness.}
\label{fig_fluxes_cor}
\end{figure}

We show as triangles on Fig. \ref{fig_fluxes_cor} the ratio of the
flux densities corrected from clustering to the flux densities
measured with PSF-fitting, as a function of $L_{\rm FUV}$. The
amplitude of the correction is blue around $-20$ per cent
  compared to the PSF-estimated flux at $L_{\rm FUV}
  <10^{10}L_{\odot}$, and becomes negligible for $L_{\rm FUV} >
  10^{10}L_{\odot}$. The amplitude of this correction is larger for
  the faint bins, where we observe a stronger departure from a PSF
  profile (see Appendix \ref{appendix_imgs}).
%Compared to a PSF-fitting flux measurement with a constant background, this correction yields a decrease of the flux by less than 20\%. Note that the effect of clustering compensates with the effect of incompleteness discussed section \ref{incomp}.  

\subsection{Summary: flux density measurements}

%obtained as described in sect. \ref{incomp},

%We summarize here the steps we follow to measure flux densities by
%stacking.

For a given stacking measure, we correct first for stacking bias by
subtracting the stacking bias profile from the stack profile, adding
errors of the profiles in quadrature. The stacking bias profile used
is derived by stacking fake sources recovered by source extraction
using the same $L_{\rm FUV}$ distribution as the galaxies in the stack
under study. We then fit the resulting profile with
eq. (\ref{eq_stack_wtheta}), leaving $\overline{S}$ and $a$ as free
parameters, in order to get a flux density corrected
  from the effect of clustering. We show in Fig. \ref{fig_fluxes_cor}
the amplitude of the correction (combining effects of incompleteness
and clustering as circles) as a function of $L_{\rm FUV}$. These
corrections partly compensate for each other; the amplitude of the
overall correction decreases with luminosity. The correction is
maximal at $L_{\rm FUV} = 6\times10^{9}\,{\rm L}_{\sun}$ and becomes
negligible for luminosities larger than $2\times10^{10}\,{\rm
  L}_{\sun}$. The overall correction is larger at
  faint luminosities because the incompleteness is more important, and
  also because of the larger departure of the observed profiles from a
  pure PSF. Note that the errors on Fig. \ref{fig_fluxes_cor} are
large for faint luminosities because the errors on $S_{\rm{PSF}}$ are
large (around twice the errors on $S_{\rm{corr}}$); the amplitude of
the correction itself is well determined.

The errors on the flux density $\overline{S}$ are obtained by boostrap
resampling, repeating the above procedure on 3000 random bootstrap
samples, and determining the error from the standard deviation of the
fluxes of these bootstrap samples.

We further performed the following tests to assess the reliability of
our results. First of all, we performed the full analysis using median
stacking, and obtained very similar results (see Appendix
\ref{appendix_median_stacking}). We also tested how detected objects
impact our stacking results. We subtracted from the SPIRE images the
sources detected at various threshold levels (from 3 to
10$\sigma$). We performed the stacking on these images, and added to
the flux measured by stacking the flux of the detected objects. The
results we obtain with this method are slightly
  higher (on average 20 per cent, see Appendix
  \ref{appendix_cleaning_stacking}) than the results presented here,
  while they agree at the 1$\sigma$ level. The results do not depend
  significantly on the threshold level we use to subtract detected
  objects. We discuss in Sect. \ref{sec_sfrd} the impact of this
  difference on the cosmic star formation rate density derived from
  the IR luminosity functions we build from the stacking
  measurements.

%Combined together, these corrections account for 20\% of the flux at most.

\section{Results}
We perform stacking at 250, 350, and 500\,$\mu$m as a function of UV
luminosity and slope of the UV continuum, $\beta$. Tables with
stacking results are in Appendix \ref{appendix_fluxes}, and we show
the postage stamp images of the stacking in Appendix
\ref{appendix_imgs}. We derive total IR luminosity, $L_{\rm{IR}}$, by
integrating over the range $8<\lambda<1000\,\mu$m of the best fit to
the \citet{Dale_2002} templates, obtained with the SED-fitting code
CIGALE \citep{Noll_2009}. There are various ways to assign a redshift
to a given stack population; we consider here as the redshift the mean
photometric redshift of the galaxies involved in the stacking. The
error on $L_{\rm{IR}}$ is given by the standard deviation of the
probability distribution function of the $L_{\rm IR}$ values obtained
with the models used during the fitting procedure
\citep[see][]{Noll_2009}.

The \citet{Dale_2002} templates are calibrated as a function of a
single parameter $\alpha$. These models assume that the dust mass over
interstellar radiation field ratio varies as a power law of the
interstellar radiation field with index $-\alpha$. We observe slight
variations of this parameter as measured by the SED fitting. As a
function of luminosity, we find that $\alpha$ decreases from $\sim
2.6$ for $L_{\rm FUV}<1.5\times 10^{10}\,{\rm L}_{\sun}$ to $\sim 1.7$
for $L_{\rm FUV}>1.5\times 10^{10}\,{\rm L}_{\sun}$, which implies an
increase of dust temperature from $\sim$ 25 to 30\,K. The significance
of this increase is however only at a level of 1.1$\sigma$. As a
function of the slope of the UV continuum, $\beta$, $\alpha$ is
roughly constant at around $1.8$, which implies a dust temperature of
around 28\,K.

%cf. fig. 6 of Chapman et al., 2002, ApJ, 588, 186
%using directly the Lir vs Tdust relation
%http://physics.uwyo.edu/~ddale/research/seds/alpha.dat for alpha - 60/100 correlation 
\subsection{Stacking as a function of UV luminosity}\label{sec_stack_Luv}
\begin{figure}
  \includegraphics[width=\hsize]{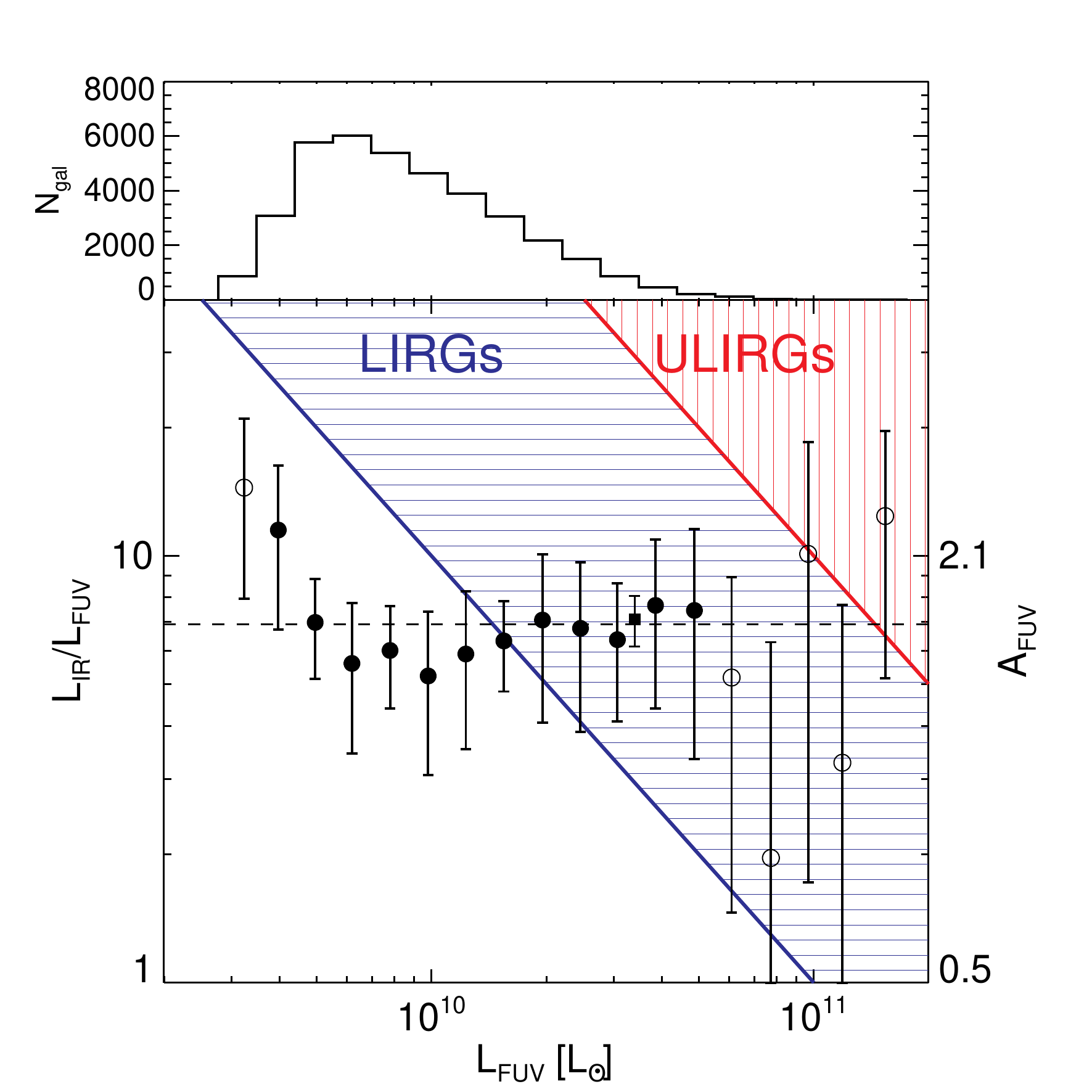}
\caption{IR-to-UV luminosity ratio as a function of UV
  luminosity. Estimates with $S/N>3$ at 250, 350 and 500\,$\mu$m
  are shown as filled circles, and others as empty circles. The
  horizontally hatched region represents the locus of LIRGs, while the
  vertically hatched region is for ULIRGs. The dashed line shows the
  mean $L_{\rm IR}/L_{\rm FUV}$ value from estimates with $S/N>3$ at
  250, 350 and 500\,$\mu$m. The filled square shows the estimate from
  stacking of UV-selected galaxies at \textit{Herschel}-PACS
  wavelengths at $z\sim2$ of \citet{Reddy_2012}, slightly offset in
  $L_{\rm FUV}$ for clarity. The right axis shows the equivalent
  attenuation in the FUV band (in magnitudes), using
  eq. \ref{eq_afuv}. The top panel shows the histogram of galaxies as
  a function of $L_{\rm FUV}$.}
\label{fig_lir_luv_luv}
\end{figure}

We stack UV-selected galaxies in bins of UV luminosity $L_{\rm FUV}$
of size 0.1\,dex. The results are presented in
Fig. \ref{fig_lir_luv_luv}\footnote{We consider here
    as signal-to-noise ratio the ratio of the flux measured after
    applying the corrections described in sect. \ref{sec_stacking} to
    the error. Note that some stacking measurements can have observed
    signal-to-noise ratios (i.e. before applying any correction) lower
    than 3.}, where we plot the ratio of IR to UV luminosities (which
is a tracer of dust attenuation) as a function of the UV
luminosity. The $L_{\rm{IR}}/L_{\rm FUV}$ ratio is found to be
constant with UV luminosity over most of the range we probe, with a
mean of $6.9\pm 1.$ This suggests that the dust
attenuation does not depend heavily on UV luminosity in a UV-selected
sample at $z\sim 1.5$ in this luminosity range. Assuming that the
relation between the attenuation in the GALEX FUV band and
$L_{\rm{IR}}/L_{\rm FUV}$ is \citep{Overzier_2011, Seibert_2005}
\begin{equation}\label{eq_afuv}
 A_{\rm FUV} = 2.5\log\left[
   \frac{1}{1.68}\left(\frac{L_{\rm{IR}}}{L_{\rm FUV}}\right) +1 \right],
 \end{equation}

\noindent the average value of the IR to UV luminosity ratio for our
sample corresponds to a value $A_{\rm FUV}=1.8 \pm
  0.1$\,mag.

%Our results, however, suggest that the $L_{\rm{IR}}/L_{\rm FUV}$ ratio
%increases by a factor of two at lower luminosities ($16.\pm8.$ at
%$L_{\rm FUV} = 3\times 10^{9} {\rm L}_{\sun}$, and $16.\pm 6.$ at
%$L_{\rm FUV} = 4\times 10^{9} {\rm L}_{\sun}$); however the
%significance of this increase is at a level of only 1.5$\sigma$. Note
%that this increase in $L_{\rm{IR}}/L_{\rm FUV}$ ratio is not an
%artefact created by the corrections presented sect. \ref{incomp}, as
%non-corrected measurements show a similar trend (see the stack images
%in \ref{appendix}).

We also show in Fig. \ref{fig_lir_luv_luv} the regions where Luminous
Infrared Galaxies (LIRGs, $10^{11}<L_{\rm{IR}}/{\rm
  L}_{\sun}<10^{12}$) and Ultra Luminous Infrared Galaxies (ULIRGs,
$L_{\rm{IR}}>10^{12}\,{\rm L}_{\sun}$) lie. Our results show that the
average IR luminosities of UV-selected galaxies in the UV luminosity
range we explore at $z\sim 1.5$ are comparable to LIRGs, but not to
ULIRGs.  Our measures of the $L_{\rm{IR}}/L_{\rm FUV}$ ratio are in
agreement with previous measures obtained from objects within
UV-selected samples detected in both the UV and IR \citep{Reddy_2006b,
  Buat_2009, Reddy_2010}.  Our result is also in excellent agreement
with the stacking study of \citet{Reddy_2012}, who found that the
average IR-to-UV luminosity ratio of a sample of 114 UV-selected
galaxies at $z\sim 2$ is $7.1\pm 1.1$, for galaxies with $L_{\rm FUV}
\sim 3.1\times 10^{10}\,{\rm L}_{\sun}$.  Note also that
\citet{Buat_2009} observed that the fraction of galaxies with
$L_{\rm{IR}}/L_{\rm FUV}>5$ is roughly constant with UV luminosity for
$L_{\rm UV}\ga 3\times 10^{9}\,{\rm L}_{\sun}$, which is consistent
with the trend we measure here.  We can also compare our result with
the average attenuation derived by \citet{Cucciati_2012} from SED
fitting of galaxies selected in the $I$-band (around
3000\,\AA~rest-frame at the mean redshift of our sample) with
$u^*g'r'i'zJHK_{\rm s}$ photometry and spectroscopic redshifts. They
derive $A_{\rm FUV} = 2.17$\,mag for the same
redshift range ($1.2<z<1.7$) as our study, which is slightly larger
than what we measure, though note that we select galaxies at shorter
restframe wavelengths.

\subsection{Stacking as a function of UV slope, $\beta$}\label{sec_stack_beta}
The slope of the UV continuum has been shown to correlate with the
dust attenuation within galaxies \citep[e.g.][]{Calzetti_1994,
  Meurer_1999}. The use of the $\beta$ slope offers an estimate of
the dust attenuation from the rest-frame UV, without requiring far-IR
data or spectral lines diagnostics. Calibrations have been derived
from spectro-photometric samples of starburst galaxies at low
redshifts \citep[e.g.][]{Meurer_1999, Overzier_2011}, and are
routinely used to derive dust attenuation at various redshifts, in
particular using slopes derived from rest-frame UV colours.

We use the $u^*$, $V$, intermediate (IA427, IA464, IA484, IA505,
IA527, IA574, IA624, IA679, IA709, IA738, IA767, IA827) and
narrow-band (NB711, NB816) filters to compute the slope $\beta$. We
adjust the photometry to a simple power-law SED, $f_{\lambda} \propto
\lambda^{\beta}$, over the restframe wavelength range
$1200<\lambda<3000$\,\AA. This means that there are at least 9 bands
available for the measure of $\beta$, and at the mean redshift of this
sample, $z=1.43$, there are 12 bands available.

\begin{figure}
\includegraphics[width=\hsize]{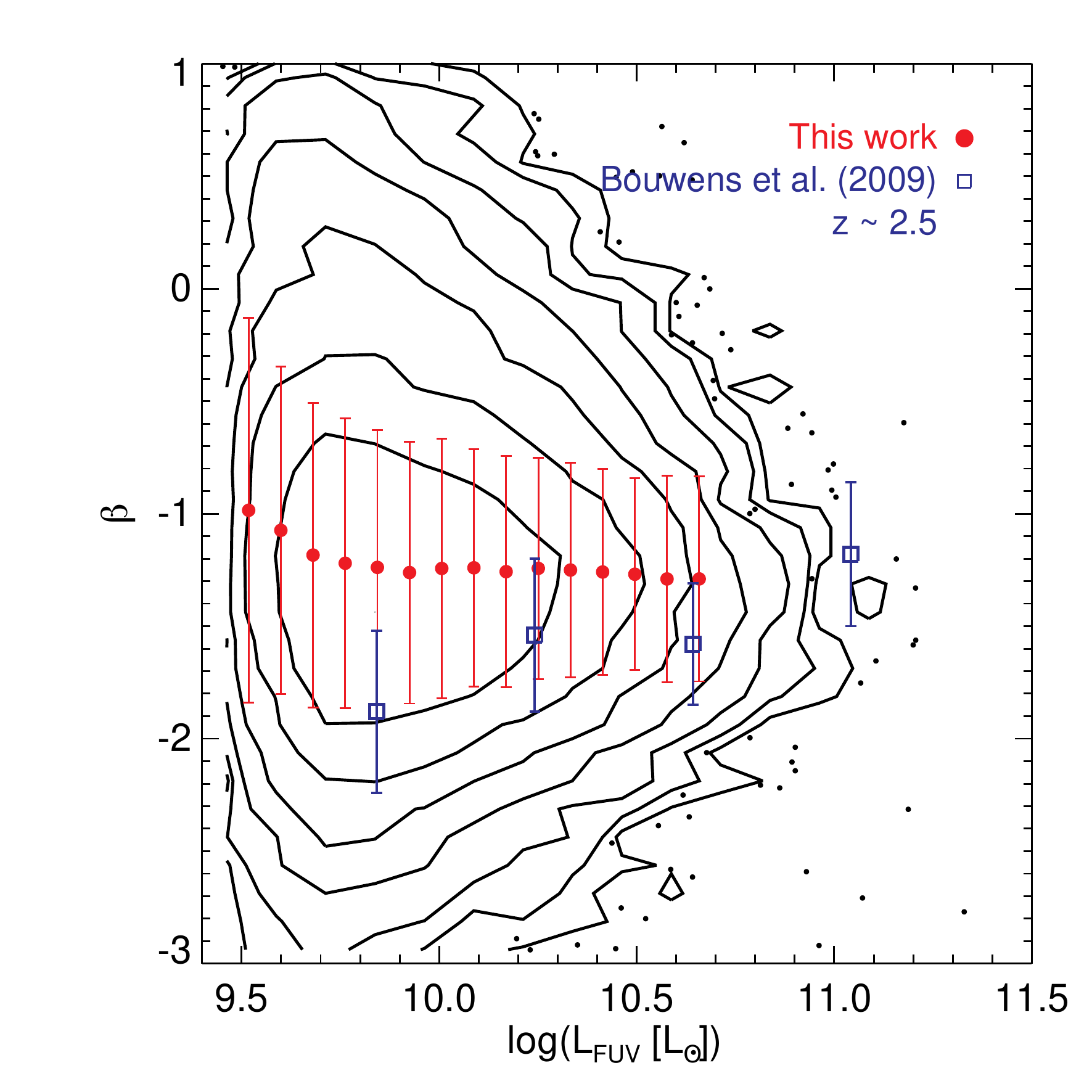}
\caption{UV continuum slope $\beta$ as a function of UV
  luminosity. The large filled circles show the mean and dispersion in
  15 bins of luminosity for $L_{\rm FUV}<5\times10^{10}\,{\rm
    L}_{\sun}$. Open squares show the mean $\beta$ measured for
  $U$-band dropouts from WFPC2 (F300W) data at $z\sim 2.5$
  \citep{Bouwens_2009}.}
\label{fig_beta_Luv}
\end{figure}

Figure \ref{fig_beta_Luv} shows the distribution of $\beta$ as a
function of UV luminosity. The mean UV slope for our sample is
$\langle \beta \rangle = -1.2 \pm 0.6$. We find that the average slope
of the UV continuum is mostly independent of the UV luminosity, while
the dispersion slightly decreases with increasing $L_{\rm FUV}$. We
compare these measures with the results of \citet{Bouwens_2009}
obtained from $U$-band dropouts from WFPC2 (F300W) data at $z\sim
2.5$. While the two distributions overlap at the $1\sigma$ level, at a
given UV luminosity the UV-selected galaxies are redder at lower
redshifts. This is in agreement with a more global trend observed from
$z\sim 7$ to $z\sim 2.5$ \citep{Bouwens_2009}. Note also that at
higher redshifts, fainter galaxies tend to be bluer \citep[with a
  significance of 5$\sigma$,][]{Bouwens_2009}, while we do not observe
such a trend at lower redshift. We note that our sample contains less
than 1 per cent of ``quiescent'' galaxies, according to the criterion
of \citet{Ilbert_2010} based on the restframe $NUV-R$ colour. Moreover
the lack of dependence of $\beta$ with UV luminosity remains if we
split galaxies in restframe $NUV-R$ colour or specific star formation
rate.
 \begin{figure}
\includegraphics[width=\hsize]{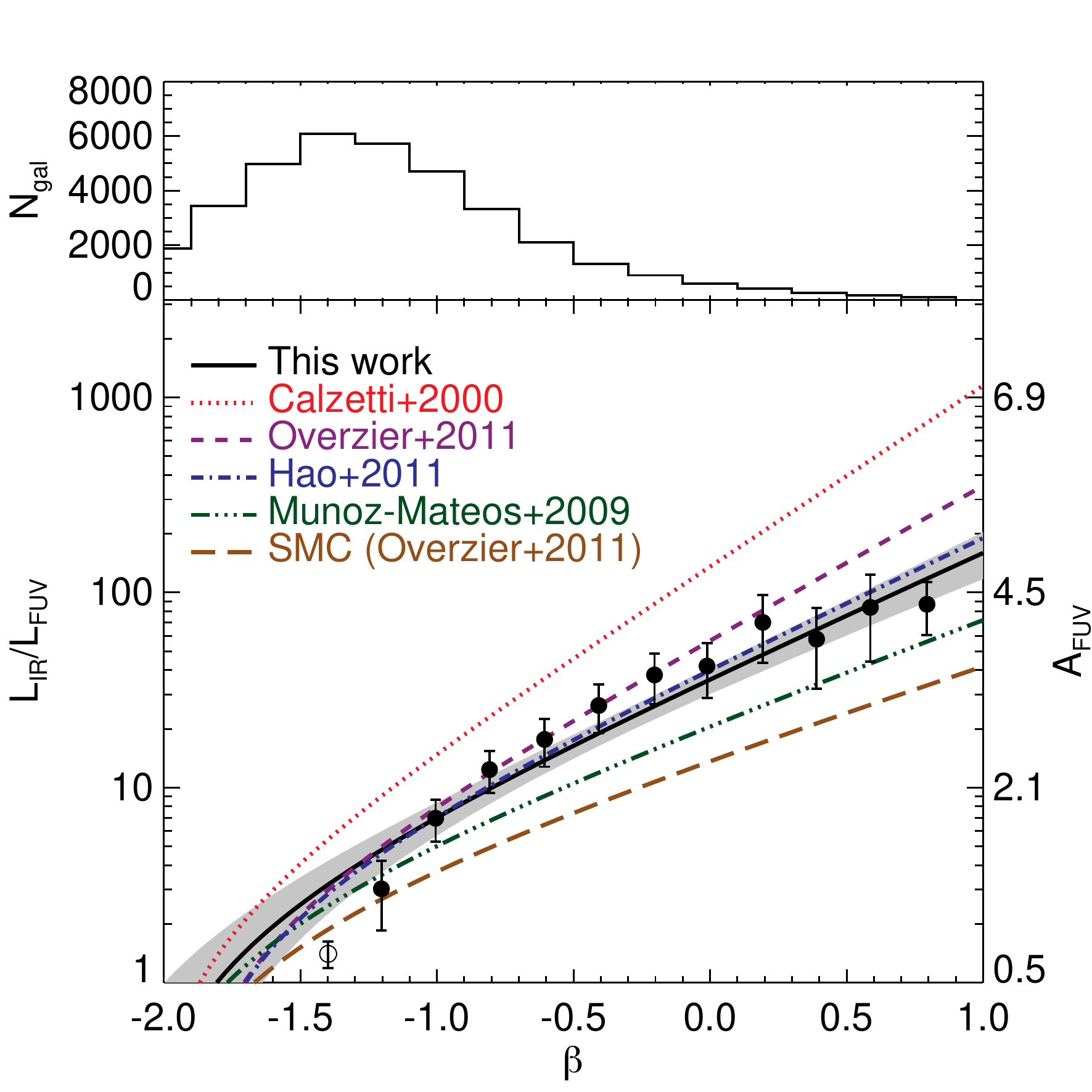}
\caption{$L_{\rm IR}/L_{\rm FUV}$ vs. $\beta$ from stacking. Estimates
  with $S/N>3$ at 250, 350 and 500\,$\mu$m are shown as filled
  circles, and others as empty circles. Lines show various $L_{\rm
    IR}/L_{\rm FUV}-\beta$ relations: \citet[][dotted]{Calzetti_2000};
  \citet[][dashed]{Overzier_2011}; \citet[][dot-dashed]{Hao_2011};
  \citet[][dot-dot-dashed]{Munoz-Mateos_2009}; and
the relation expected for the SMC extinction curve
    \citep[from][long dashed]{Overzier_2011}. The solid line shows
  the best fit to our measurements (considering only estimates with
  $S/N>3$), and the grey area the range of relations implied by the
  1$\sigma$ errors on the parameters. The top panel shows the number
  of galaxies in each bin of $\beta$.}
\label{fig_irx_beta}
\end{figure}
 
We show on Fig. \ref{fig_irx_beta} the IR-to-UV luminosity ratio as a
function of $\beta$. Note that the stacking results do not cover the
same range in $\beta$ as in Fig. \ref{fig_beta_Luv}: the signal is
small for $\beta <-1.5$.

%for $\beta <-2$ the signal is small, as well as for $\beta = -1.6$.

We observe a good correlation between the UV slope and dust
attenuation. We compare our results to several calibrations of the
$(L_{\rm IR}/L_{\rm FUV})--\beta$ relation derived for local starburst
galaxies \citep[][their `total relation']{Calzetti_2000,
  Overzier_2011}. The difference between these two local relations
comes from the fact that \citet{Overzier_2011} remeasured the UV
photometry for the \citet{Meurer_1999} sample, hence possibly
including fewer starburst regions.

%In the case of \citet{Calzetti_2001}, we use the relation
 
% \begin{equation}\label{eq_irx}
% \log\left[ \frac{1}{1.68}\left(\frac{L_{IR}}{L_{0.16}}\right) +1 \right] = \frac{A_{FUV}}{2.5} = 0.84(\beta - \beta_0)
% \end{equation}
%where $\beta_0$ here is taken to be equal to -2.1, which is the value
%expected for a continuous star formation over 1Gyr
%\citep{Calzetti_2000}. Considering shorter timescales would imply a
%higher $L_{TIR}/L_{FUV}$ for a given value of $\beta$.
 
Our results fall between the local starburst and the local calibration
for normal star-forming galaxies from \citet{Munoz-Mateos_2009} and the relation expected for the Small Magellanic Cloud
  (SMC) extinction curve. Our results are, however, consistent with
the relation obtained by \citet{Hao_2011} from another set of local
normal star-forming galaxies, based on SINGS observations
\citep[][]{Kennicutt_2003}, as well as data from
\citet{Moustakas_2006}. Note that the spread between the calibrations
from \citet{Munoz-Mateos_2009} and \citet{Hao_2011} comes from
differences in galaxy selections.

Following previous work we fit our measurements assuming
\begin{equation}\label{eq_irx}
 A_{\rm FUV} = 2.5\log\left[
   \frac{1}{1.68}\left(\frac{L_{\rm{IR}}}{L_{\rm FUV}}\right) +1 \right] =
 a+b\beta.
 \end{equation}

We consider only measurements with $S/N>3$ in the fit; including other
measurements does not affect the results. We find $a = 3.4\pm 0.1$ and
$b=1.6\pm0.1$. Our value for the slope $b=dA_{\rm FUV}/d\beta$ of this
relation is lower than the values derived from commonly used local
starburst attenuation laws (e.g. \citet{Meurer_1999} find $b = 1.99$;
\citet{Calzetti_2000} find $b = 2.31$), but larger than the value derived at intermediate redshifts
($1<z<2$) by \citet{Buat_2011}, namely $b = 1.46\pm 0.21$. The implied
UV slope for the dust-free case is $\beta_0 = -2.12\pm0.18$; this is
in agreement with what is expected from stellar population models
\citep{Leitherer_1995}, and favours a continuous star-formation mode.

%. Note however that $\beta_0$ is related to the age of the stellar
%populations dominating the UV emission
%\citep[e.g.][]{Calzetti_2001,Leitherer_1995}, and also that a small
%error on $\beta_0$ has a significant impact on the derived amount of
%dust extinction. For instance an error of 0.2 in $\beta_0$ implies in
%our case an error of 30\% on the estimated $L_{FIR}/L_{FUV}$.
  
\section{UV and IR luminosity functions}\label{sec_LF}
In this section, we present our determination of the UV luminosity
function of our UV sample. Using the stacking results presented above,
we can also recover the total far-IR luminosity function of this
sample. This procedure enables us, for instance, to discuss the
dust-corrected contribution to the star-formation rate density of
UV-selected galaxies.

\citet{Reddy_2008} performed a similar study on a sample of LBGs at
$1.9<z<3.4$. They derived the $E(B-V)$ distribution of their sample by
maximising the likelihood of observing their data for a given
luminosity, redshift and reddening distribution. Assuming the
\citet{Meurer_1999} attenuation law, they determined the IR luminosity
function of UV-selected galaxies using a Monte Carlo method and found
that at $z\sim 2$ it is in agreement with the IR luminosity function
of 8\,$\mu$m rest-frame selected galaxies with luminosities
$10^{10}\,{\rm L}_{\sun} - 10^{12}\,{\rm L}_{\sun}$ range. This
contrasts with the lower redshift result of \citet{Buat_2009}, who
directly measured the IR luminosity function of UV-selected galaxies
at $z\sim 1$ and noticed that it underestimates the IR luminosity
function of 12\,$\mu$m rest-frame selected galaxies for
$L_{\rm{IR}}\ga2\times10^{11}\,{\rm L}_{\sun}$.

We derive luminosity functions using the $V_{\rm max}$ method
\citep{Schmidt_1968}. In practice, we derive for each galaxy of our
sample the minimum ($z_{\rm min}$) and maximum ($z_{\rm max}$)
redshifts where it can be included in the sample given its redshift
and luminosity: $z_{\rm min} = \max(1.2, z_{\textrm{UV}, \rm min})$,
and $z_{\rm max} = \min(1.7, z_{\textrm{UV},\rm
  max})$. $z_{\textrm{UV}, \rm min}$ and $z_{\textrm{UV},\rm max}$ are
the minimum and maximum redshifts implied by the magnitude limits we
used to build our UV-selected sample.

The maximum volume within which this galaxy can be observed is then
given by:

\begin{eqnarray}
  V_{\rm max}  & = & V(z_{\rm max}) - V(z_{\rm min}) \\
  V(z)    & = & \frac{A}{3} r_{\rm c}^3(z), \nonumber
\end{eqnarray}
where $A$ is the solid angle covered by the observations, and $r_{\rm
  c}(z)$ is the comoving distance. We corrected for incompleteness as
a function of luminosity using the simulations described
Sect. \ref{sec_data}. We define our completeness limit as the
luminosity where the incompleteness is equal to 20 per cent; this
corresponds to $L_{\rm FUV} = 5\times 10^9 \rm{L}_{\sun}$ and $L_{\rm
  IR} = 5.6\times10^{10}\rm{L}_{\sun} $. We also include the error on
the completeness correction in the luminosity function errors. For
each luminosity function estimate, we take into account the error on
photometric redshifts by constructing 50 mock catalogues with new
redshifts within the probability distribution functions derived by
\citet{Ilbert_2009}. Note that this procedure yields an estimate of
the errors added by the use of photometric redshifts, but not of the
bias they introduce.

%We consider as completeness limit the lowest luminosity where the
%incompleteness is lower than 15 per cent

 \begin{figure}
\includegraphics[width=\hsize]{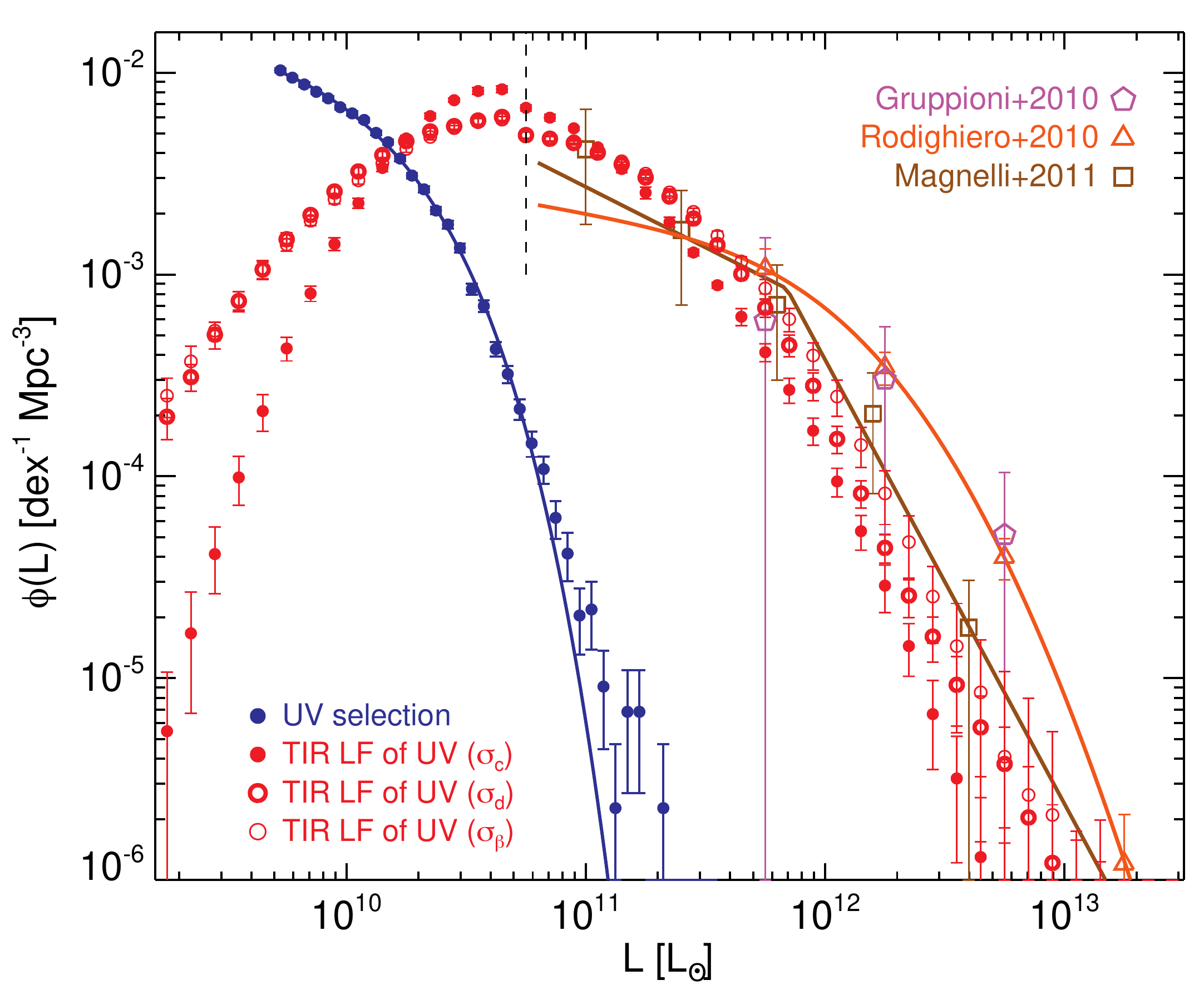}
\caption{UV and IR luminosity functions. The blue circles represent
  the UV luminosity function of our sample. The red circles show the
  IR luminosity function of this sample derived from the stacking
  measurement of the IR to UV luminosity ratio: filled circles using a
  constant dispersion of $\log(L_{\rm{IR}}/L_{\rm FUV})$ (method
  `$\sigma_c$'); open thick circles the dispersion which reproduces
  the observed $\log(L_{\rm{IR}}/L_{\rm FUV})$ values for detected
  objects (`$\sigma_d$'); and open thin circles the dispersion derived
  from the dispersion in $\beta$ slope (`$\sigma_{\beta}$'). The
  dashed line shows the lower luminosity limit that we consider to
  adjust the IR luminosity function.  Open squares show the IR
  luminosity function of an IR selected sample at $1.3<z<1.8$ from
  \citet{Magnelli_2011}; open triangles the IR luminosity function
  from \citet{Rodighiero_2010} at $1.2<z<1.7$; and open hexagons the
  IR luminosity function from \textit{Herschel}/PACS data at
  $1.2<z<1.7$ \citep{Gruppioni_2010}.}
\label{fig_lfs}
\end{figure}

\subsection{UV luminosity function}\label{sec_UV_lf}
%Once again, we derived here the rest-frame UV luminosities directly
%from the observed $u^*$-band fluxes.

We checked that the K-corrections are minimal and do not have an
impact on the UV luminosity function. We note that the faint-end of
the UV LF is quite sensitive to the method used for the
photometry. The UV LF derived using the $u^*$-band fluxes from the
catalog of \citet{Capak_2007} is significantly steeper ($3\sigma$ significance level, $\alpha =
  -1.29\pm0.03$) at the faint end than the
  measurement obtained using the photometry we use
  here. \citet{Capak_2007} performed the source extraction in a
combined $i^+$ and $i^*$ image, ran PSF matching to the image with
worst seeing ($K_s$), and finally measured aperture photometry and
aperture corrections. We attribute this difference between the LFs to
an overestimation of the $u^{*}$ flux because of a combination of the
PSF matching and the aperture corrections. We show our UV LF in Fig.
\ref{fig_lfs} (blue circles). The error bars on the UV LF are the
combination of the Poisson error, the error on the completeness
correction, and the standard deviation of the estimates from the mock
catalogues used to compute the LF. We fit the UV luminosity function
with a Schechter form; we find $\log(\phi^{\ast}
  [\textrm{Mpc}^{-3}]) = -2.24\pm 0.02$, $\log(L^{\ast} [{\rm
      L}_{\sun}]) = 10.13\pm0.02$ (equivalent to $M^* =
  -19.44\pm0.04$), and $\alpha = -1.13 \pm 0.04$. We note that the
luminosity function we derive is quite flat with respect to a number
of previous studies devoted to estimating the UV luminosity functions
at similar redshifts \citep{Arnouts_2005, Oesch_2010}, who report
$-1.6<\alpha<-1.4$. This difference could be due to the fact that
these estimates were obtained from selections using shorter restframe
wavelengths than ours. However, our estimate is similar to a recent
and independent derivation of the UV luminosity function by
\citet{Cucciati_2012}, who found $\alpha=-1.09\pm0.23$ in the same
redshift range, using spectroscopic data.

\subsection{Recovering the IR luminosity function}\label{sec_IR_lf}

From the stacking results presented above, we recover the IR
luminosity function of our UV-selected sample. To do so, we assign an
IR luminosity to each of the galaxies of our sample using the
following method, which is similar to that used by \citet{Reddy_2008}.

We assume that at a given UV luminosity, the distribution of $\log
(L_{\rm{IR}}/L_{\rm FUV})$ is Gaussian. We use as the mean of this
distribution the values of $L_{\rm{IR}}/L_{\rm FUV}$ derived from the
stacking analysis. In the luminosity range where we do not have
reliable stacking measurements ($L_{\rm FUV} >5\times 10^{10}\,{\rm
  L}_{\sun}$), we assume that the $L_{\rm{IR}}/L_{\rm FUV}$ ratio is
constant, and equal to 6.8, the average value of the ratio over
$10^{10}<L_{\rm FUV}/{\rm L}_{\sun} <5\times
10^{10}$. We also consider as mean of the
  $L_{\rm{IR}}/L_{\rm FUV}$ distribution the results obtained from an
  alternative stacking method, excluding sources detected at
  $10\sigma$ from the SPIRE images (see Appendix
  \ref{appendix_cleaning_stacking}).

%we adjust in $L_{\rm FUV}$ bins of 0.1 dex size the standard deviation
%of the Gaussian distribution such that the distribution of the
%$L_{\rm{IR}}/L_{\rm FUV}$ ratio we generate is consistent with that
%for objects detected at \textit{Herschel}-SPIRE wavelengths

%Alternatively, (\textit{ii}) using as mean of the Gaussian
%distribution the values of $L_{\rm{IR}}/L_{\rm FUV}$ derived from the
%stacking analysis, we adjust in the same $L_{\rm FUV}$ bins the
%standard deviation of the Gaussian distribution such that the
%distribution of the $L_{\rm{IR}}/L_{\rm FUV}$ ratio we generate
%reproduces the ratios of objects detected at \textit{Herschel}-SPIRE
%wavelengths.

 %We then use as the mean of the Gaussian distribution the values of
 %$L_{\rm{IR}}/L_{\rm FUV}$ derived from the stacking analysis
 %(Sect. \ref{sec_stack_Luv}), and adjust in the same $L_{\rm FUV}$
 %bins the standard deviation of the Gaussian such that the
 %distribution of the $L_{\rm{IR}}/L_{\rm FUV}$ ratio we generate
 %reproduces the observed ratios.

For the standard deviation of the Gaussian distribution, we use three
different methods. Firstly, (\textit{i}) we assume that the dispersion
of the Gaussian distribution is constant with UV luminosity and equal
to 0.35. This value has been derived at low redshifts from UV-selected
samples by \citet{Buat_2009}; we call this method
`$\sigma_{\rm{c}}$'. Alternatively, (\textit{ii}), we use as reference
the values of the $L_{\rm{IR}}/L_{\rm FUV}$ ratios of the objects
detected at \textit{Herschel}-SPIRE wavelengths (see
Sect. \ref{sec_data}); these objects represent the upper tail of the
distribution of the $L_{\rm{IR}}/L_{\rm FUV}$
ratio. We determined the mean of this distribution
  from stacking as a function of $L_{\rm FUV}$
(Sect. \ref{sec_stack_Luv}). We assume once again that the
distribution of the $L_{\rm{IR}}/L_{\rm FUV}$ ratio is Gaussian. Then
for each $L_{\rm FUV}$ stacking bin, we adjust the standard deviation
of the Gaussian such that we reproduce the observed distribution,
given the mean and the few detections in the upper tail. The resulting
dispersion is a function of $L_{\rm FUV}$, and decreases from 0.5 at
$L_{\rm FUV}=3\times10^9 {\rm L}_{\sun}$ to 0.2 at $L_{\rm
  FUV}=10^{11}{\rm L}_{\sun}$ (method `$\sigma_{\rm d}$'). Finally,
(\textit{iii}) we use the observed dispersion of the slope of the UV
continuum, $\beta$, as a function of UV luminosity, and translate it
into a dispersion in $\log (L_{\rm{IR}}/L_{\rm FUV})$, using the
relation we derived in Sect. \ref{sec_stack_beta}, $A_{\rm FUV} = 3.4
+ 1.6\beta$. This also yields a function of $L_{\rm FUV}$; the
resulting dispersion is slightly higher than the one obtained with
scenario $\sigma_{\rm d}$, decreasing from 0.6 at $L_{\rm
  FUV}=3\times10^9 {\rm L}_{\sun}$ to 0.3 at $L_{\rm FUV}=10^{11}{\rm
  L}_{\sun}$ (method `$\sigma_{\beta}$').

For a given galaxy in our sample, we then randomly assign a value of
$\log (L_{\rm{IR}}/L_{\rm FUV})$, following the relevant distribution,
whose mean and standard deviation are determined by the UV luminosity
of the galaxy. We can then derive the IR luminosity for each galaxy in
our sample, and, using the $V_{\rm max}$ values determined according
to the UV selection, compute the IR LF of the sample.

At $z=0$, all UV-selected galaxies are detected in the IR, and the
distribution of $\log (L_{\rm{IR}}/L_{\rm FUV})$ is well described by
a Gaussian \citep{Buat_2009}. Note that the actual distribution of
$\log (L_{\rm{IR}}/L_{\rm FUV})$ at $z\sim 1.5$ is not known. Assuming
a Gaussian distribution enables us to compare with previous studies
similar to ours \citep[e.g.][]{Reddy_2008}. However, depending on the
value of the dispersion of this Gaussian, a significant number of
galaxies with low $L_{\rm FUV}$ can be assigned high $L_{\rm{IR}}$.

%For each iteration we also create a mock catalogue, taking into
%account the error on the photometric redshift by assigning new
%redshifts to the objects according to their PDFs, as performed for the
%computation of the UV luminosity function.

We generate 100 realisations of this IR LF, and show on
Fig. \ref{fig_lfs} the mean and errors of these iterations (red
circles).  We compare our result with the IR LF derived at $1.3<z<1.8$
by \citet{Magnelli_2011} from a sample of galaxies selected at
24\,$\mu$m and also using stacking at 70\,$\mu$m (open squares), the
IR LF of \citet{Rodighiero_2010} (open triangles), from a sample of
galaxies at $1.2<z<1.7$, also selected at 24\,$\mu$m with mid-IR data,
and finally the \textit{Herschel}/PACS-derived IR LF from
\citet{Gruppioni_2010} at $1.2<z<1.7$ (open
hexagons)\footnote{We show in an Appendix
    (Fig. \ref{fig_lf_cleaning_stacking}) the IR LF obtained using the
    alternate stacking measurements.}.

Our results show that correcting a UV-selected sample for dust enables
us to recover the IR LF at the faint luminosities reached by the IR
selections ($L_{\rm{IR}} < 3\times 10^{11}\,{\rm L}_{\sun}$); our
estimates are in agreement with the results of \citet{Magnelli_2011}
at these luminosities.

At IR luminosities brighter than $L_{\rm{IR}} = 3\times 10^{11}\,{\rm
  L}_{\sun}$, the results depend on the assumptions about the shape of
the distribution of the $L_{\rm{IR}}/L_{\rm FUV}$ ratio. A higher
dispersion in this ratio yields a higher amplitude of the LF at the
bright end. We note however that for the scenario with the highest
dispersion, $\sigma_{\beta}$, it is not clear that the dispersion in
$\beta$ is completely related to the dispersion in dust
attenuation. First of all, we do not observe a stacking detection at
$S/N>3$ in all \textit{Herschel}-SPIRE bands for $\beta <-1.2$. Note
also that the dispersion in $L_{\rm{IR}}/L_{\rm FUV}$ that we infer
from the dispersion in $\beta$ is larger than the dispersion we derive
from the objects directly detected by SPIRE. In any case, none of the
scenarios we explore here for the dispersion of this ratio is able to
reproduce accurately the bright end of the IR selected LFs. This is
implied directly by the stacking results (see
Fig. \ref{fig_lir_luv_luv}), which show that the IR luminosities of
our sample galaxies are consistent with LIRGs, but not ULIRGs.

Hereafter, we consider as our best estimate the IR LF determined with
the scenario $\sigma_{\rm{d}}$ described above.

%Hereafter we consider as a synthetic IR LF the mean of the IR LFs
%determined with the scenarios $\sigma_{\rm{c}}$ and $\sigma_{\rm{d}}$
%described above, and as error bars the quadrature sum of the errors on
%the individual IR LFs.
%
% This synthetic LF is represented as filled circles in
% Fig. \ref{fig_uv_faint}.

\subsection{Implications for Cosmic star-formation density estimation}\label{sec_sfrd}

\begin{table*}
\begin{minipage}{100mm}
\caption{LF fit parameters and derived Cosmic Star-Formation rate density}
\label{tab_lfs}
\begin{tabular}{@{}lcccc}
\hline
\multicolumn{2}{l}{IR LF from UV, DPL fit}&\multicolumn{1}{c}{ }&\multicolumn{2}{c}{IR LF from UV, DE fit}\\
\hline
$\log(\phi_{\rm knee})$ &$-2.70\pm0.02$ & & $\log(\phi^{\ast})$ & $-2.31\pm 0.04$\\
$\log(L_{\rm knee})$ & $11.49\pm0.02$ & & $\log(L^{\ast})$ & $11.10\pm 0.09$\\
$\alpha_2$ & $-2.10\pm0.08$ & & $\sigma$& $0.41\pm0.03$\\
$\alpha_1$ & $-0.6$ & & $\alpha$ & 1.2\\
$\dot{\rho}_{\ast,\rm IR}\quad(10^{11}<L_{\rm{IR}}/{\rm L}_{\sun}<10^{13})$ & 0.08$\pm 0.01$ & &  & 0.08$\pm 0.02$\\
$\dot{\rho}_{\ast,\rm IR}\quad (10^{7\phantom{0}}<L_{\rm{IR}}/{\rm L}_{\sun}<10^{15})$ & 0.16$\pm0.01$& &  & 0.13$\pm 0.03$\\
\hline
\end{tabular}
%\medskip
DPL stands for Double Power Law, and DE for Double
Exponential. $\alpha_1$ and $\alpha$ are kept fixed during the fitting
procedure. $\phi_{\rm knee}$ and $\phi^{\ast}$ are given in Mpc$^{-3}$,
$L_{\rm knee}$ and $L^{\ast}$ in ${\rm L}_{\sun}$ and $\dot{\rho}_{\ast,\rm IR}$ in
$\rm{M}_{\sun} \rm{yr}^{-1}$Mpc$^{-3}$.
\end{minipage}
\end{table*}

We study here the implications of our results for the estimation of
the cosmic star-formation rate density (SFRD) from UV-selected samples.

We compute the UV luminosity density using $\dot{\rho}_{*,\rm UV} =
\phi_{\ast} L_{\ast} \Gamma(\alpha +2)$, and then convert it to a star
formation rate density using the relation from \citet{Kennicutt_1998}:

%the current data do not allow us to constrain the faint end slope of the LF, so we fix its value to $\alpha = -1.5$. This value is consistent with what is measured from UV selected samples at $0.8<z<3.5$ \citep{Arnouts_2005}.   

\begin{equation}
SFR [\rm{M}_{\sun} \rm{yr}^{-1}] = 1.4\times10^{-28} L_{\nu} [\rm{erg}\; \rm{s}^{-1}\; \rm{Hz}^{-1}],
\end{equation}
which assumes a \citet{Salpeter_1955} Initial Mass Function (IMF).  We
find a star-formation density $\dot{\rho}_{\ast, \rm UV} = 0.022 \pm
0.001 \, \rm{M}_{\sun} \rm{yr}^{-1}$Mpc$^{-3}$; the error on
$\dot{\rho}_{\ast, \rm UV}$ is derived from the allowed range of UV
LFs within $1\sigma$ of the best fit.

IR luminosity function is commonly described by a double power law
\citep[DPL, ][]{Sanders_2003}:
\begin{eqnarray}
	\phi(L) &=& \phi^{\ast}\left(\frac{L}{L_{\rm knee}}\right)^{\alpha_1} \quad L<L_{\rm knee} \nonumber;\\
	\phi(L) &=& \phi^{\ast}\left(\frac{L}{L_{\rm knee}}\right)^{\alpha_2} \quad L>L_{\rm knee}.
\end{eqnarray}

Alternatively one can use a double exponential
\citep[DE,][]{Saunders_1990}:

\begin{equation}
\phi(L) = \phi^* \left(\frac{L}{L^*}\right)^{1-\alpha}\exp\left[-\frac{1}{\sigma^2}\log^2\left(1+\frac{L}{L^*}\right)\right].
\end{equation} 

%We adjust our synthetic IR LF using
%$L_{\rm{IR}}=6.5\times10^{10}\,{\rm L}_{\sun}$ as a completeness limit
%(dashed line on Fig. \ref{fig_lfs}).

We use both parameterisations, in order to compare with the results of
\citet{Magnelli_2011}, who use the DPL form, and
\citet{Rodighiero_2010} who use the DE form. We allow to be free
parameters the normalisations, $\phi_{\rm knee}$ and $\phi^{\ast}$,
the characteristic luminosities, $L_{\rm knee}$ and $L^{\ast}$, and
the bright-end slopes, $\alpha_2$ and $\sigma$.  We keep the faint-end
slopes fixed, using the same assumptions as \citet{Magnelli_2011}
($\alpha_1 = -0.6$) and \citet{Rodighiero_2010} ($\alpha = 1.2$).  The
best-fit parameters are given in Table \ref{tab_lfs} for our baseline IR LF.
  
We then integrate the IR LF within two luminosity ranges:
$10^{7}<L_{\rm{IR}}/{\rm L}_{\sun}<10^{15}$, which implies an
extrapolation of the measured luminosity function to faint
luminosities, and $10^{11}<L_{\rm{IR}}/{\rm L}_{\sun}<10^{13}$, which
is the range where observations are available. We finally convert the
IR luminosity density to a SFRD using
\begin{equation}
SFR [\rm{M}_{\sun} \rm{yr}^{-1}] = 4.5\times10^{-44} L_{\rm{IR}} [\rm{erg}\;\rm{s}^{-1}]
\end{equation}
\citep{Kennicutt_1998}, which also assumes a Salpeter IMF.
The resulting star-formation rate densities are listed in Table \ref{tab_lfs}.

\begin{figure}
\includegraphics[width=\hsize]{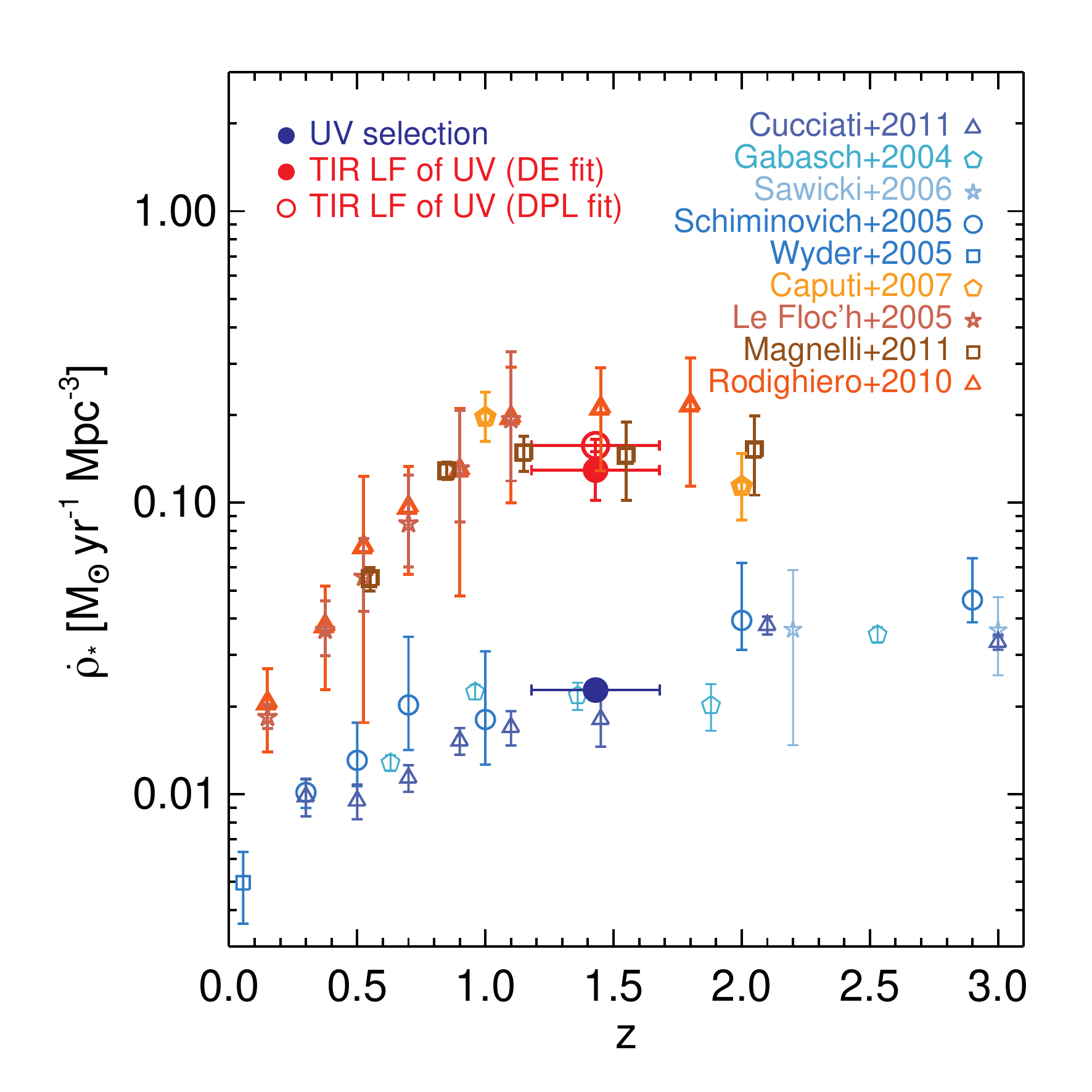}
\caption{Cosmic star-formation rate history. Our results are
  represented as filled circles: UV selection (blue); and IR LF of the
  UV selection (red). We compare these results to the estimates based
  on UV-selected samples (blue-like colours): \citet[open
    hexagons,][]{Gabasch_2004}; \citet[open
    circles,][]{Schiminovich_2005}; \citet[open
    square,][]{Wyder_2005}; \citet[open stars,][]{Sawicki_2006}; and
  \citet[open triangles,][]{Cucciati_2012}. We also show estimates
  based on IR selections (red-like colours): \citet[][open
    hexagons]{Lefloch_2005}; \citet[][open hexagons]{Caputi_2007};
  \citet[][open triangles]{Rodighiero_2010}; and \citet[][open
    squares]{Magnelli_2011}.}
\label{fig_sfrd}
\end{figure}

%Note, however, that this measure involves an extrapolation of the IR
%LF to faint luminosities.

We consider first our most secure estimate of the SFRD, from the
luminosity range where measurements are available
($10^{11}<L_{\rm{IR}}/{\rm L}_{\sun} <10^{13}$). The SFRD derived from
our baseline IR LF of the UV-selected galaxies is
about $89\pm 32$ per cent of the density derived
from the result of \citet{Magnelli_2011}, and $52 \pm
  25$ per cent of the density derived from \citet{Rodighiero_2010}
from IR-selected samples. If we consider the IR LF
  built from the alternative stacking results, these percentages are
  $112 \pm 41$ and $65\pm37$ per cent. 

%This is a direct consequence of the fact that the IR LF of the UV
%selection \textcolor{blue}{might lack} a significant contribution from
%ULIRGs.

We show our results for the cosmic star-formation rate extrapolating
the luminosity function to low luminosity in Fig. \ref{fig_sfrd} along
with other UV-selected and IR-selected measurements (references on the
figure), all integrated over the same luminosity ranges, and converted
to the same IMF.  The result from our UV selection is in good
agreement with previous determinations of the star-formation rate
density from UV-selected samples.

The cosmic SFR we derive from the IR LF of the UV selection is in
broad agreement with the estimates from IR-selected samples at similar
redshifts. Nevertheless, given the differences in the estimates of the
SFRD from IR-selected samples, the percentage of the IR SFRD recovered
by our method varies from $100 \pm 33$ per cent if
we consider \citet{Magnelli_2011} to $61 \pm 27$ per cent considering
\citet{Rodighiero_2010}. If we consider the IR LF
  built from the alternative stacking method, these percentages are
  $131 \pm40$ and $76 \pm 38$ per cent. This suggests that, after
  correction for dust attenuation, a UV-selected sample at $z\sim1.5$
  down to $L_{\rm FUV}=3\times 10^{9}\,\rm{L}_{\sun}$ can recover the
  total SFRD estimated from the IR, the remaining uncertainty being on
  the discrepancy between the IR-selected LFs.

%This suggests that, even after correction for dust attenuation, a
%UV-selected sample at $z\sim1.5$ down to $L_{\rm FUV}=3\times
%10^{9}\,\rm{L}_{\sun}$ may miss a significant fraction of the SFRD.

Our results also imply that the dust-corrected estimate of the SFR
density is roughly 6 times higher than the determination from direct
UV observations.
 
We can compare our SFRD estimate with what would be derived using an
average correction derived from $\beta$. Specifically, we derive the
average attenuation factor $(\rm{SFR}_{\rm IR}+\rm{SFR}_{\rm
  UV})/\rm{SFR}_{\rm UV} \simeq 0.64 L_{\rm IR}/L_{\rm FUV}+1$ from
the distribution of $\beta$, following the same method as
\citet{Bouwens_2012}. We use our best fit for the relation between
$A_{\rm FUV}$ and $\beta$ (see Sect. \ref{sec_stack_beta}). We can
then determine the UV SFRD obtained from the UV LF integrated over the
full data range (i.e. down to $0.2\times L_{*,\rm FUV}$), and correct
it for dust attenuation with this average factor. We compare this
value to the SFRD derived from the integration of the IR LF of the UV
selection over the full IR luminosity range we probe. The UV SFRD
corrected for dust attenuation using $\beta$ is around 25 per cent lower than the SFRD derived from the IR
LF of the UV selection. This suggests that, at $z=1.5$ at least, using
an average dust correction factor derived from $\beta$ can lead to a
significant underestimation of the SFRD.

%\textbf{To do}
% \begin{itemize}
% \item Add discussion about the required amount of dust extinction needed to ma%tch the bright end of the TIR LF
% \item Test standard deviation of $L_{TIR}/L_{UV}$ ratio (can find dispersion i%n order to match the IR selected derived LF ?)
 %\item Discuss the comparison with \citet{Reddy_2008}
 % \end{itemize}

\section{Discussion}\label{sec_discussion}

\subsection{Comparison with previous studies}

There have been a number of studies exploring similar topics at lower
and higher redshifts than our selection. We compare these with our
results for the relation between the slope of the UV continuum and the
dust attenuation, as well as the measure of the IR LF of a UV-selected
sample.

\subsubsection{Relation $\beta$ slope - $L_{\rm{IR}}/L_{\rm FUV}$}\label{disc_beta}
We find that there is a correlation between the slope of the UV
continuum and the ratio $L_{\rm{IR}}/L_{\rm FUV}$. However, the relation
we observe at $z=1.5$ is different from the relation that is derived
from calibrations performed at low redshifts on starburst galaxies
\citep{Meurer_1999, Calzetti_2000}. This starburst relation is
commonly used over a wide redshift range to correct UV luminosities
for dust attenuation when far-IR measurements are not available
\citep[e.g.][]{Adelberger_2000, Schiminovich_2005, Bouwens_2009}.

It is claimed that the \citet{Meurer_1999} and \citet{Calzetti_2000}
relations are valid at various redshifts: \citet{Reddy_2006a} show
that typical $z\sim2$ UV-selected galaxies detected at 24\,$\mu$m do
follow the Meurer et al. relation. \citet{Magdis_2010a} and
\citet{Magdis_2010b} also find that the UV star-formation rates of
Lyman Break Galaxies at $z\sim3$ corrected for attenuation with the
\citet{Meurer_1999} relation are in agreement with far-IR and radio
estimates.  \citet{Reddy_2006a} find, however, that the dust
attenuation is overestimated for galaxies with younger stellar
populations; they also show that galaxies with lower SFRs tend to lie
under the \citet{Meurer_1999} relation. \citet{Burgarella_2007} notice
that UV star-formation rates corrected for dust attenuation with the
Meurer et al. relation are in agreement with IR-derived ones, but that
the dispersion is large. In a study similar to ours,
\citet{Reddy_2012} perform stacking of $z\sim2$ Lyman Break Galaxies
at 24, 100 and 160\,$\mu$m, as well as 1.4\,GHz. They find that the
average $L_{\rm{IR}}/L_{\rm FUV}$ ratio of these $L_{\ast}$ Lyman
Break Galaxies is consistent with the local starburst relation. 

%We note that in addition to the results presented in
%Sect. \ref{sec_stack_beta} we re-measured $\beta$ slopes using the
%same method as \citet{Reddy_2012}, by fitting a simple power-law SED,
%$f_{\lambda} \propto \lambda^{\beta}$, to the $u^*$ and $V_J$
%photometry, and found no impact on our results.

While our result hence might seem at odds with previous work, a number
of studies have shown that attempting to correct for dust attenuation
using the slope of the UV continuum is quite complex, and requires
some caution.

At low redshift, \citet{Bell_2002}, \citet{Boissier_2007},
\citet{Munoz-Mateos_2009}, and \citet{Seibert_2005} showed that the
$\beta$ slope and the dust attenuation of normal star-forming galaxies
do not follow the same relationship as for local starbursts; in this
case the \citet{Meurer_1999} relation overpredicts the dust
attenuation, which is in agreement with our results. The validity of
the starburst relation depends also on the sample selection: for local
galaxies dust attenuation might be overestimated for a UV-selected
sample, while underestimated for IR-selected samples \citep{Buat_2005,
  Seibert_2005}.  On the other hand, at higher redshifts
($0.66<z<2.6$), \citet{Murphy_2011} observed that the
\citet{Meurer_1999} relation overpredicts the dust attenuation for
24\,$\mu$m selected galaxies. Note also that \citet{Buat_2010} found
that at $z<0.3$, galaxies selected at 250\,$\mu$m have dust attenuations
between the local starbursts and the normal star forming relation from
\citet{Boissier_2007}. The mean relation that we derive here is
actually in excellent agreement with their results.  We note also that
our sample is dominated by galaxies of IR luminosities similar to
LIRGs. At $z\sim1.5$, these galaxies belong mostly to the `main
sequence' of star forming galaxies and are not in the starburst mode
\citep{Sargent_2012}, which can partly explain why the relation
between $\beta$ and the dust attenuation we obtain is similar to
normal star forming galaxies at low redshifts.

The variety of results presented above is linked to the fact that the
UV and IR emissions can come from different regions in the galaxies,
and also that attempting to use the slope of the UV continuum to
correct for dust attenuation requires assumptions about the underlying
extinction law and star-formation history. Using the
\citet{Meurer_1999} or the \citet{Calzetti_2000} relations is
consistent with using quite shallow extinction laws, such as the
\citet{Calzetti_1994} or \citet{Calzetti_1997} ones, which is
equivalent to adopting a clumpy foreground distribution of dust. Our
results (in terms of the slope of the $A_{\rm FUV}-\beta$ relation)
suggest a steeper extinction law, which is expected for dust
geometries similar to a foreground screen. Note
  however that using the SMC extinction law yields a relation lower
  than our results, and hence is too steep.  On the other hand, the
value of the dust-free slope we derive suggests that the star
formation mode is more continuous rather than starburst.

In summary, our result for the $\beta$ slope - $L_{\rm{IR}}/L_{\rm
  FUV}$ relation at $=1.5$ is similar to those obtained from local
normal star forming galaxies. This suggests that our sample galaxies
have a steeper extinction law than those from \citet{Calzetti_1994}
and \citet{Calzetti_1997}, but less steep than the
  SMC one. This also shows that using a $\beta$ slope -
$L_{\rm{IR}}/L_{\rm FUV}$ relation calibrated on local starburts
galaxies can induce an overestimation of the $L_{\rm{IR}}/L_{\rm FUV}$
from the $\beta$ slope by around a factor of 2.

%This is equivalent to make a number of assumptions about the relative
%distribution of the dust and the star forming regions. %on one hand,
%and about star formation history on the other hand.  The relatively
%shallow \citet{Calzetti_1994} extinction law is indeed expected in
%the case of a foreground clumpy distribution of dust. Our results
%suggest a steeper extinction law, which is expected for dust
%geometries similar to foreground screen. %Finally,

%Several previous studies actually pointed out that the
%\citet{Meurer_1999} law might not be valid for the overall star
%forming galaxy population at low and high redshift

%While of course the \citet{Meurer_1999} law is an average relation,
%and some dispersion around it is expected, one must be cautious when
%using it on a given sample. 

\subsubsection{IR LF of UV-selected sample}
We find that the IR LF of a UV-selected sample at $z\sim 1.5$ down to
$L_{\rm FUV}=3\times10^{9}\,\rm{L}_{\odot}$ recovers the faint-end of
the IR LF of far-IR selected samples, but might
underestimate the bright-end, if we consider the
  latest \textit{Herschel} results, and our most conservative stacking
  measures.

These results are in agreement with those of
\citet{Buat_2009} ($z\sim 1$) who found that the bolometric ($L=L_{\rm
  UV}+L_{\rm{IR}}$) LF of a UV selection directly measured from UV and
IR data underestimates the IR LF from IR selected sample for $L>
2\times 10^{11}\,{\rm L}_{\sun}$. At higher redshifts,
\citet{Reddy_2008} performed a similar study, and found, on the
contrary, that the reconstructed IR LF of UV-selected samples at $z=2$
and $z=3$ is similar to the IR LF of IR-selected samples. Note that we
use a method which is quite similar to that of \citet{Reddy_2008} to
build our IR LF. In detail, \citet{Reddy_2008} reconstruct two IR LFs:
one from the distribution of $E(B-V)$ derived from a maximum
likelihood analysis; and the other one using previously observed
$L_{\rm{IR}}/L_{\rm FUV}$ ratios and dispersion. These two LFs are
consistent with each other. We note that their method of using the
distribution of $E(B-V)$ assumes the \citet{Calzetti_2000} relation,
which does not seem to apply to our sample (Sect. \ref{disc_beta}). In
other words, since using the \citet{Calzetti_2000} relation on our
sample would yield larger dust-corrected luminosities, this can
explain part of the discrepancy.

The other method used by \citet{Reddy_2008} is based on previous
determinations of the IR-to-UV ratio and its dispersion
\citep[$L_{\rm{IR}}/L_{\rm FUV} = 4.7$,
  $\sigma(\log(L_{\rm{IR}}/L_{\rm FUV})) =
  0.53$,][]{Reddy_2006a}. This average IR-to-UV ratio is lower than
the values we derive here, while the dispersion is higher. In order to
test the impact of our assumptions on the recovery of the bright-end
of the IR LF, we used as an extreme case the values of
$L_{\rm{IR}}/L_{\rm FUV}$ we obtained here, and the dispersion used by
\citet{Reddy_2008} to construct the IR LF from the UV selection. The
results we thus obtain (see filled squares on
fig. \ref{fig_faint_end}) are in agreement with the IR LF of
\citet[][based on \textit{Spitzer} data]{Magnelli_2011}, but then
slightly underestimate the measurements of \citet[][also based on
  \textit{Spitzer} data]{Rodighiero_2010} and \citet[][based on
  \textit{Herschel} PACS data]{Gruppioni_2010}. While this result is
in better agreement with the LFs obtained from IR-selected samples, we
believe that a constant dispersion of the $(L_{\rm{IR}}/L_{\rm FUV})$
with $L_{\rm FUV}$ is unlikely. Based on our sample, we observe that
the dispersion around $(L_{\rm{IR}}/L_{\rm FUV})$ decreases with
$L_{\rm FUV}$, either from the dispersion in $\beta$ or from the
dispersion which is required to reproduce the $(L_{\rm{IR}}/L_{\rm
  FUV})$ values of the few UV-selected galaxies detected at the SPIRE
wavelengths. Note also that using this value of the dispersion would
imply, for instance, around four times as many ULIRGs as we detect
with SPIRE. We also attempted to adjust the dispersion required to
match the IR LF of \citet{Rodighiero_2010}, assuming an exponentialy
decreasing function of $L_{\rm FUV}$, as we observe in the data. The
best fit in this case implies twice as many ULIRGs as we detect.

\subsubsection{Contribution to the CIB of UV-selected sources}
Our results for the IR LF of our UV-selected sample imply that a
UV-selection misses a significant part of the IR galaxy population, at
least within the UV luminosity range we are able to probe. To
investigate this further, we measured the contribution to the Cosmic
Infrared Background (CIB) at the SPIRE wavelengths from the
UV-selected galaxies in our sample, and compared it to the values
measured by \citet{Bethermin_2012}, who derived the contribution to
the CIB from 24\,$\mu$m-selected sources with $S_{24}>80\,\mu$Jy. We
measured by stacking the average flux density for galaxies in two
redshift bins ($1.2<z<1.4$ and $1.4<z<1.6$), and converted it to a
surface brightness. We find that the contribution to the CIB from our
UV-selected galaxies is lower than that from 24$\mu$m-selected
sources. The contribution to the CIB from our UV-selected galaxies is
around 50 per cent of that from 24\,$\mu$m-selected sources for
$1.2<z<1.4$ and around 30 per cent for $1.4<z<1.6$. This result
clearly shows that a UV selection is missing a part of the galaxy
population probed by IR selections, and that the amount of energy
which is emitted by this missing population is a significant fraction
of the CIB.

\subsection{Recovering the bright-end of the IR LF}\label{sel_disc_bright_end}

\begin{figure}
\includegraphics[width=\hsize]{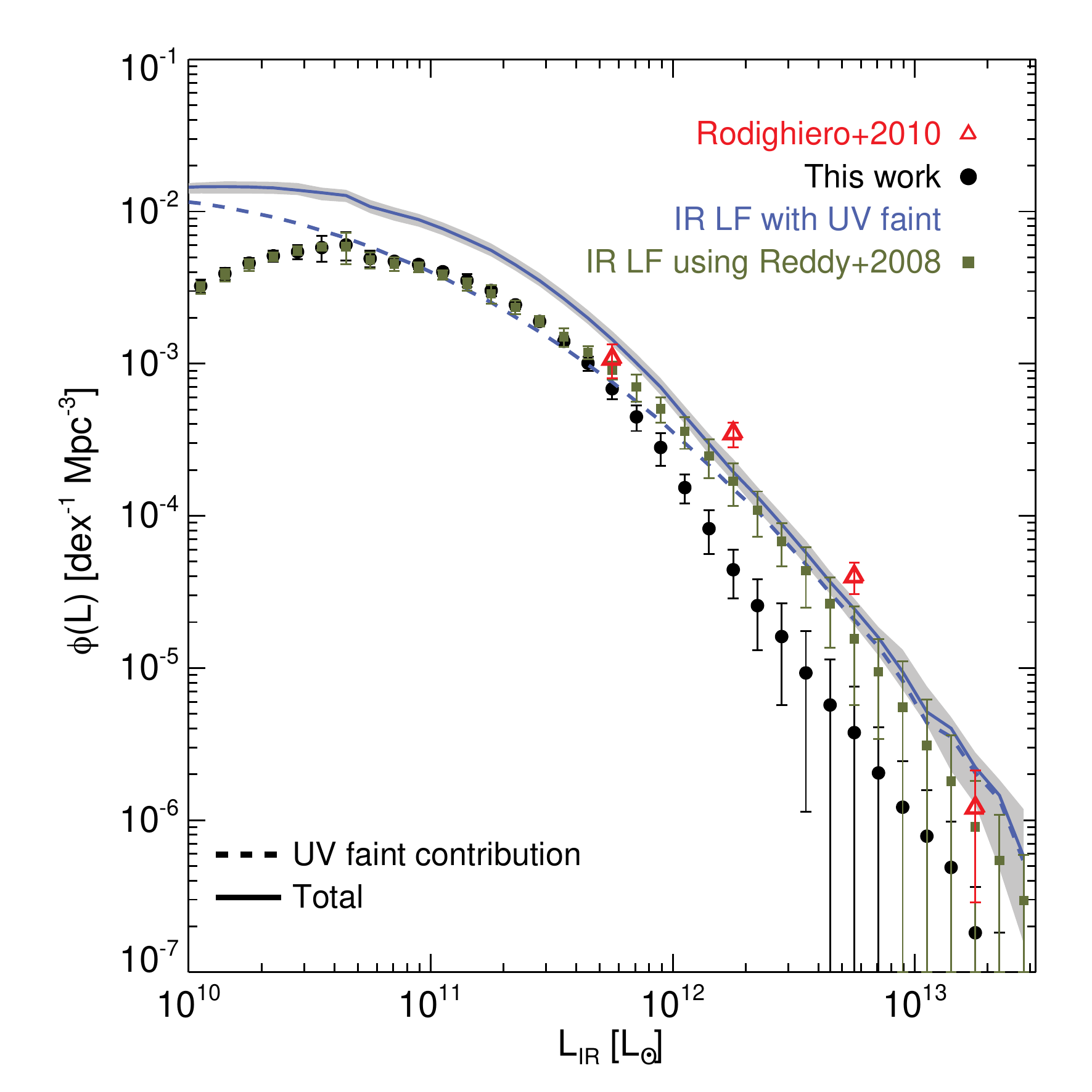}
\caption{Recovering the bright-end of the IR LF. Filled circles show
  the IR LF of our UV selection, based on available data. The solid
  line show the total IR LF of the UV selection including the
  contribution of fainter galaxies ($10^8<L_{\rm FUV}/{\rm
    L}_{\sun}<10^{9.5}$) adjusted to match the result of
  \citet{Rodighiero_2010}. The dashed line show the contribution to
  the IR LF of the faint UV galaxies.  The shaded grey area shows the
  range of LFs implied by the $1\sigma$ errors on the parameters. We
  assume here that the dispersion of the ratio
  $\log(L_{\rm{IR}}/L_{\rm FUV})$ is a constant for galaxies with
  $10^8<L_{\rm FUV}/{\rm L}_{\sun}<10^{9.5}$. The filled squares show
  the IR LF of the UV selection (no extrapolation to luminosities
  fainter than our limit) we obtain if we use our stacking
  measurements for $\log(L_{\rm{IR}}/L_{\rm FUV})$ and the dispersion
  on this ratio used by \citet{Reddy_2008}.}
\label{fig_faint_end}
\end{figure}

Using our most conservative stacking measurements,
  our results show that the IR LF we build for our UV-selected sample
does not recover the bright-end of the measured IR LF from IR-selected
samples, if we consider the latest \textit{Herschel}
  estimates. This would imply that a part of the IR
galaxy population is missed by UV selection, at least in the redshift
and luminosity ranges we consider here. We investigate the possibility
that this missing population is actually fainter in UV luminosity than
the limit reached by our observations. Note that according to our
results shown in Fig. \ref{fig_lir_luv_luv}, the galaxies in our
UV-selected sample are mostly similar to LIRGs, hence we have to
populate the ULIRGs regime of the luminosity function, which requires
a large dispersion of the $L_{\rm{IR}}/L_{\rm FUV}$ ratio.

To do so, we extrapolate the UV luminosity function down to $L_{\rm
  FUV} = 10^{8}\,{\rm L}_{\sun}$, using the best fit we obtain (see
Sect. \ref{sec_UV_lf}).  This corresponds to a magnitude limit
$u^*=29.75$\,mag, 3.75 magnitudes deeper than the
data we are using. In practice we create a mock catalogue which has
the proper luminosity function. Then we compute the IR LF of this mock
catalogue using the same method described in Sect. \ref{sec_IR_lf}. In
this case, however, we do not have measurements for the mean IR-to-UV
ratio or the dispersion of this ratio. We rather adjust these two
quantities such that the sum of the IR LF of the mock catalogue and
the IR LF of the UV selection from our baseline
  stacking method is consistent with the IR LF of
\citet{Rodighiero_2010}.

%(\textit{i}) a constant (we note this scenario as '$\sigma_{\rm c}$');
%(\textit{ii}) we extrapolate at $10^8 < L_{\rm FUV}/L_{\sun} < 4\times
%10^{9}$ the observed dispersion on the beta slope (scenario noted
%'$\sigma_{\beta}$')

%\footnote{We found that the dispersion of the IR to UV luminosity
%  ratio implied by the dispersion of the $\beta$ slope can be
%  approximated by $\sigma\left(\log(L_{\rm{IR}}/L_{\rm FUV})\right) =
%  0.80\exp^{-\left(\log(L_{\rm FUV})-9.04\right)}$.}.

As we are limited by the relatively poor observational constraints on
the IR LF, we make the following simplistic assumptions: we consider
that the ratio $L_{\rm{IR}}/L_{\rm FUV}$ and the dispersion around
this ratio are constant over the range $10^8 < L_{\rm FUV}/{\rm
  L}_{\sun} < 3\times10^{9}$; we then use two free parameters,
$\langle L_{\rm{IR}}/L_{\rm FUV} \rangle$ and
$\sigma\left(\log(L_{\rm{IR}}/L_{\rm FUV})\right)$, to adjust the
contribution of UV faint galaxies to the bright-end of the IR LF.

The results of the fit are shown Fig. \ref{fig_faint_end}. The
contribution to the IR LF of the faint UV galaxies is shown as the
dashed line, and the total IR LF as the solid line. These results show
that it is possible to obtain parameters that fit the IR LF measured
from IR-selected samples. We find $\langle L_{\rm{IR}}/L_{\rm FUV}
\rangle=8.7^{+3.1}_{-2.7}$ and $\sigma(\log(L_{\rm{IR}}/L_{\rm FUV}))
= 0.73^{+0.04}_{-0.06}$. The average IR to UV ratio required to match
the bright-end of the IR LF is consistent with the value we measure in
our two faintest bins at $L_{\rm FUV} = (3-4)\times 10^{9}\,{\rm
  L}_{\sun}$ ($L_{\rm{IR}}/L_{\rm FUV} = 14.5\pm6, 11.5\pm5$), while
the requested dispersion of the ratio is larger than what we observe:
at $L_{\rm FUV} = 3\times 10^{9}\,{\rm L}_{\sun}$ the dispersion
required to match the detected objects is
$\sigma(\log(L_{\rm{IR}}/L_{\rm FUV})) = 0.5$.

%On the other hand, if we consider scenario $\sigma_{\beta}$, the
%average IR-to-UV ratio requested is much lower. This is a consequence
%of the fact that the extrapolation to faint luminosities of the
%observed dispersion of $\beta$ yields quite large values in the range
%we explore here.

From this fit we can estimate the fraction of galaxies that is missed
by a UV selection, compared to an IR selection. We integrate the IR LF
of the faint UV galaxies in the range $6.5\times10^{10} <
L_{\rm{IR}}/{\rm L}_{\sun} < 10^{13}$, where data are available, and compare
this density to the one derived by integrating the total IR LF in the
same range. We find that in terms of number density, 56 per cent of galaxies
are missed.

The relatively poor constraints on the IR LF do not enable us to draw
firm conclusions on the scenarios of dust attenuation for faint UV
galaxies. Note that the shape of the IR LF is quite sensitive to the
assumptions on the dispersion $\sigma(\log(L_{\rm{IR}}/L_{\rm
  FUV}))$. Hence we expect to set better constraints on these
scenarios with updated \textit{Herschel} LFs.

%There is, however, indication that this ratio might increase towards
%lower UV luminosities ($L_{\rm FUV}<4\times10^{9} {\rm L}_{\sun}$).

\section{Summary and Conclusions}\label{sec_conclusion}

We used \textit{Herschel}-SPIRE \citep{Griffin_2010,Swinyard_2010}
imaging from the HerMES \citep{Oliver_2012} programme to study the IR
properties of a sample of UV-selected galaxies at $z\sim 1.5$ in the
COSMOS field. We built our sample from galaxies detected in CFHT $u^*$
band \citep{Capak_2007} down to $u^*=26$\,mag, with
photometric redshifts \citep{Ilbert_2009} $1.2<z_{\rm phot}<1.7$. Only
a few per cent of these galaxies are detected at the
\textit{Herschel}-SPIRE wavelengths, so we use stacking in order to
derive the average IR luminosities of the galaxies as a function of
$L_{\rm FUV}$ and the slope of the UV continuum $\beta$. We detail the
techniques we used to correct the stacking measurements from stacking
bias and clustering of the UV-selected galaxies, based on extensive
simulations. We use these stacking measurements to derive the IR LF of
the UV-selected sample, in order to infer the cosmic star formation
rate density probed by a UV selection at $z=1.5$. Our conclusions are
as follows:

\begin{enumerate}[1.]
\item UV-selected galaxies at $z=1.5$ and
  $4\times10^{9}<L_{\rm FUV}/\rm{L}_{\sun}<5\times10^{10}$ have average total
  IR luminosities similar to LIRGs, but not to ULIRGs.
\item The average $L_{\rm{IR}}/L_{\rm FUV}$ ratio is roughly constant,
  with $L_{\rm FUV}$ ($4\times10^{9}<L_{\rm FUV}/{\rm L}_{\sun} <5
  \times 10^{10}$) and is equal to $6.9 \pm 1.$
\item The average $L_{\rm{IR}}/L_{\rm FUV}$ ratio is correlated with
  the slope of the UV continuum, $\beta$. This relation is below the
  relation derived from local starburst galaxies, but is in agreement
  with previous results obtained from local normal star-forming
  galaxies. Our best fit to this relation is $A_{\rm FUV} = 3.4\pm0.1
  + (1.6\pm0.1)\beta$.
\item We built the IR LF of the UV sample using our stacking
  measurements of the average $L_{\rm IR}/L_{\rm FUV}$ ratio, and
  assuming that the distribution of $\log(L_{\rm IR}/L_{\rm FUV})$ is
  Gaussian. We used three different scenarios for the value of the
  dispersion $\sigma(\log(L_{\rm IR}/L_{\rm FUV}))$, which all yield
  the same result that the IR LF of the UV sample is in reasonable
  agreement at the faint-end ($L_{\rm{IR}} \sim 10^{11}\,{\rm
    L}_{\sun}$) with the IR LF from IR-selected samples at the same
  epoch, but might underestimate it at the
  bright-end ($L_{\rm{IR}} \ga 5\times10^{11}\,{\rm L}_{\sun}$).
\item At $z\sim1.5$ a UV rest-frame selection without dust attenuation
  correction probes roughly 10 per cent of the total (UV+IR)
  star-formation rate density. The cosmic star-formation rate density
  derived from the IR LF of the UV sample corresponds to
  61--76 per cent or 100--133 per
  cent of the star-formation rate density derived from IR-selected
  samples, depending on the IR LF taken as reference
  (\citealt{Rodighiero_2010} or \citealt{Magnelli_2011}).
\item Assuming our most conservative measures and the
  latest \textit{Herschel} estimates, the fraction of galaxies which
  are missed by a UV selection compared to an IR selection at $z\sim
  1.5$ is around 50 per cent, in terms of number density; this number
  is sensitive to the assumptions on the dispersion
  $\sigma(\log(L_{\rm IR}/L_{\rm FUV}))$.
\end{enumerate}

\section*{Acknowledgements}
We thank the referee for a careful reading and detailed, constructive
comments which helped improving the paper. S.H. and V.B. thank the
French Space Agency (CNES) for financial support. SPIRE has been
developed by a consortium of institutes led by Cardiff Univ. (UK) and
including: Univ. Lethbridge (Canada); NAOC (China); CEA, LAM (France);
IFSI, Univ. Padua (Italy); IAC (Spain); Stockholm Observatory
(Sweden); Imperial College London, RAL, UCL-MSSL, UKATC, Univ. Sussex
(UK); and Caltech, JPL, NHSC, Univ. Colorado (USA). This development
has been supported by national funding agencies: CSA (Canada); NAOC
(China); CEA, CNES, CNRS (France); ASI (Italy); MCINN (Spain); SNSB
(Sweden); STFC, UKSA (UK); and NASA (USA).

We thank the COSMOS collaboration for sharing data used in this paper.

The data presented in this paper will be released through the {\em
  Herschel} Database in Marseille (HeDaM,
{http://hedam.oamp.fr/HerMES})

\appendix
\section{Stacking results}\label{appendix_fluxes}
\begin{table*}
\begin{minipage}{130mm}
\caption{Stacking results as a function of $L_{\rm FUV}$}
\label{tab_stack_lfuv}
\begin{tabular}{@{}cccccccc}
\hline
$\log(L_{\rm FUV} [L_{\odot}])$ range & $\langle \log(L_{\rm FUV} [L_{\odot}]) \rangle$ & $\langle z\rangle$ & $N_{gal}$ & $S_{250}$ [mJy] & $S_{350}$ [mJy] & $S_{500}$ [mJy] & $\log(L_{\rm IR} [L_{\odot}])$ \\
\hline
9.44 -- 9.54 & 9.51 & 1.26 & 872 & 0.71 $\pm$ 0.24 & 0.84 $\pm$ 0.21 & 0.81 $\pm$ 0.27 & 10.67 $\pm$ 0.20\\
9.54 -- 9.64 & 9.60 & 1.34 & 3075 & 0.66 $\pm$ 0.12 & 0.72 $\pm$ 0.16 & 0.52 $\pm$ 0.15 & 10.66 $\pm$ 0.18\\
9.64 -- 9.74 & 9.70 & 1.43 & 5760 & 0.46 $\pm$ 0.08 & 0.58 $\pm$ 0.10 & 0.44 $\pm$ 0.10 & 10.54 $\pm$ 0.11\\
9.74 -- 9.84 & 9.79 & 1.46 & 6018 & 0.43 $\pm$ 0.10 & 0.53 $\pm$ 0.10 & 0.36 $\pm$ 0.09 & 10.54 $\pm$ 0.17\\
9.84 -- 9.94 & 9.89 & 1.44 & 5383 & 0.63 $\pm$ 0.09 & 0.69 $\pm$ 0.11 & 0.54 $\pm$ 0.10 & 10.67 $\pm$ 0.12\\
9.94 -- 10.04 & 9.99 & 1.44 & 4631 & 0.65 $\pm$ 0.10 & 0.66 $\pm$ 0.12 & 0.48 $\pm$ 0.10 & 10.71 $\pm$ 0.18\\
10.04 -- 10.14 & 10.09 & 1.44 & 3883 & 0.88 $\pm$ 0.11 & 0.89 $\pm$ 0.13 & 0.59 $\pm$ 0.11 & 10.86 $\pm$ 0.17\\
10.14 -- 10.24 & 10.19 & 1.43 & 3052 & 1.35 $\pm$ 0.13 & 1.40 $\pm$ 0.16 & 1.02 $\pm$ 0.13 & 10.99 $\pm$ 0.10\\
10.24 -- 10.34 & 10.29 & 1.43 & 2177 & 1.54 $\pm$ 0.17 & 1.42 $\pm$ 0.16 & 0.90 $\pm$ 0.14 & 11.14 $\pm$ 0.18\\
10.34 -- 10.44 & 10.39 & 1.43 & 1503 & 1.87 $\pm$ 0.21 & 1.79 $\pm$ 0.20 & 1.10 $\pm$ 0.19 & 11.22 $\pm$ 0.19\\
10.44 -- 10.54 & 10.49 & 1.44 & 873 & 2.43 $\pm$ 0.25 & 2.38 $\pm$ 0.29 & 1.60 $\pm$ 0.26 & 11.29 $\pm$ 0.15\\
10.54 -- 10.64 & 10.59 & 1.44 & 463 & 3.45 $\pm$ 0.40 & 3.32 $\pm$ 0.43 & 2.10 $\pm$ 0.34 & 11.47 $\pm$ 0.19\\
10.64 -- 10.74 & 10.69 & 1.44 & 211 & 3.66 $\pm$ 0.61 & 3.39 $\pm$ 0.61 & 1.99 $\pm$ 0.54 & 11.56 $\pm$ 0.24\\
10.74 -- 10.84 & 10.78 & 1.43 & 111 & 3.56 $\pm$ 0.77 & 3.09 $\pm$ 0.85 & 2.42 $\pm$ 0.85 & 11.50 $\pm$ 0.31\\
10.84 -- 10.94 & 10.89 & 1.45 & 35 & 3.85 $\pm$ 1.62 & 2.46 $\pm$ 1.92 & 0.00 $\pm$ 0.75 & 11.18 $\pm$ 0.96\\
10.94 -- 11.04 & 10.99 & 1.43 & 16 & 9.50 $\pm$ 1.58 & 7.50 $\pm$ 1.97 & 4.69 $\pm$ 1.92 & 11.99 $\pm$ 0.36\\
11.04 -- 11.14 & 11.07 & 1.52 & 5 & 3.97 $\pm$ 2.36 & 3.89 $\pm$ 3.51 & 4.50 $\pm$ 4.69 & 11.59 $\pm$ 0.58\\
11.14 -- 11.24 & 11.19 & 1.44 & 5 & 22.96 $\pm$ 8.25 & 22.23 $\pm$ 7.21 & 17.63 $\pm$ 5.04 & 12.28 $\pm$ 0.25\\
\hline
\end{tabular}
\medskip
\end{minipage}
\end{table*}

\begin{table*}
\begin{minipage}{130mm}
\caption{Stacking results as a function of $\beta$}
\label{tab_stack_beta}
\begin{tabular}{@{}ccccccccc}
\hline
$\beta$ range & $\langle \beta \rangle$ & $\langle \log(L_{\rm FUV} [L_{\odot}])\rangle$ & $\langle z\rangle$ & $N_{gal}$ & $S_{250}$ [mJy] & $S_{350}$ [mJy] & $S_{500}$ [mJy] & $\log(L_{\rm IR} [L_{\odot}])$ \\
\hline
-2.50 -- -2.30 & -2.39 & 9.92 & 1.46 & 418 & 0.00 $\pm$ 0.03 & 0.08 $\pm$ 0.17 & 0.17 $\pm$ 0.23 & 8.46 $\pm$ 0.72\\
-2.30 -- -2.10 & -2.19 & 9.98 & 1.46 & 921 & 0.00 $\pm$ 0.00 & 0.00 $\pm$ 0.00 & 0.00 $\pm$ 0.00 & 0.00 $\pm$ 0.00\\
-2.10 -- -1.90 & -1.99 & 9.98 & 1.45 & 1877 & 0.00 $\pm$ 0.00 & 0.00 $\pm$ 0.00 & 0.00 $\pm$ 0.00 & 0.00 $\pm$ 0.00\\
-1.90 -- -1.70 & -1.79 & 10.04 & 1.45 & 3448 & 0.00 $\pm$ 0.00 & 0.00 $\pm$ 0.00 & 0.00 $\pm$ 0.00 & 0.00 $\pm$ 0.00\\
-1.70 -- -1.50 & -1.60 & 10.05 & 1.45 & 4984 & 0.00 $\pm$ 0.00 & 0.00 $\pm$ 0.00 & 0.03 $\pm$ 0.04 & 6.25 $\pm$ 0.66\\
-1.50 -- -1.30 & -1.40 & 10.07 & 1.44 & 6081 & 0.19 $\pm$ 0.02 & 0.38 $\pm$ 0.09 & 0.28 $\pm$ 0.10 & 10.22 $\pm$ 0.07\\
-1.30 -- -1.10 & -1.20 & 10.07 & 1.43 & 5725 & 0.44 $\pm$ 0.09 & 0.52 $\pm$ 0.11 & 0.38 $\pm$ 0.11 & 10.55 $\pm$ 0.17\\
-1.10 -- -0.90 & -1.00 & 10.05 & 1.42 & 4705 & 1.11 $\pm$ 0.10 & 1.09 $\pm$ 0.11 & 0.84 $\pm$ 0.10 & 10.89 $\pm$ 0.10\\
-0.90 -- -0.70 & -0.81 & 10.02 & 1.41 & 3321 & 1.76 $\pm$ 0.13 & 1.79 $\pm$ 0.12 & 1.18 $\pm$ 0.10 & 11.11 $\pm$ 0.11\\
-0.70 -- -0.50 & -0.61 & 9.99 & 1.41 & 2109 & 2.31 $\pm$ 0.17 & 2.22 $\pm$ 0.17 & 1.44 $\pm$ 0.15 & 11.24 $\pm$ 0.12\\
-0.50 -- -0.30 & -0.41 & 9.99 & 1.40 & 1325 & 3.40 $\pm$ 0.22 & 3.08 $\pm$ 0.24 & 2.09 $\pm$ 0.20 & 11.41 $\pm$ 0.12\\
-0.30 -- -0.10 & -0.20 & 9.96 & 1.39 & 899 & 4.59 $\pm$ 0.30 & 4.10 $\pm$ 0.32 & 2.72 $\pm$ 0.27 & 11.54 $\pm$ 0.13\\
-0.10 -- 0.10 & -0.01 & 9.94 & 1.38 & 604 & 4.88 $\pm$ 0.38 & 4.48 $\pm$ 0.40 & 2.88 $\pm$ 0.33 & 11.56 $\pm$ 0.14\\
0.10 -- 0.30 & 0.19 & 9.87 & 1.37 & 411 & 6.38 $\pm$ 0.49 & 5.57 $\pm$ 0.45 & 3.42 $\pm$ 0.43 & 11.72 $\pm$ 0.16\\
0.30 -- 0.50 & 0.39 & 9.87 & 1.36 & 266 & 6.18 $\pm$ 0.79 & 5.71 $\pm$ 0.71 & 3.74 $\pm$ 0.54 & 11.63 $\pm$ 0.19\\
0.50 -- 0.70 & 0.59 & 9.83 & 1.36 & 164 & 6.26 $\pm$ 0.59 & 5.48 $\pm$ 0.65 & 3.17 $\pm$ 0.59 & 11.75 $\pm$ 0.20\\
0.70 -- 0.90 & 0.79 & 9.77 & 1.34 & 107 & 8.44 $\pm$ 1.05 & 8.68 $\pm$ 0.99 & 5.69 $\pm$ 0.72 & 11.71 $\pm$ 0.13\\
\hline
\end{tabular}
\medskip
\end{minipage}
\end{table*}

\section{Stacking postage stamp images}\label{appendix_imgs}

%We show here the stacked images used for the analysis; 

We display here only the stacking results with signal-to-noise ratios
larger than 3 at 250, 350 and 500\,$\mu$m. This signal-to-noise ratio
is the ratio of the flux density over the error on the flux density
after the corrections described in Sect.~\ref{sec_stacking} have been
applied. We show here the raw stacking images, before applying any
correction. All stack images are 240\,arcsec accross. The gray-scale
shows the signal, using a asinh stretch \citep{Lupton_2004}. Each
panel uses a different scale, driven by the maximum flux. We also show
for reference the radial profile along with the fit by the proper PSF
in each band.
\begin{figure*}
\includegraphics[width=\hsize]{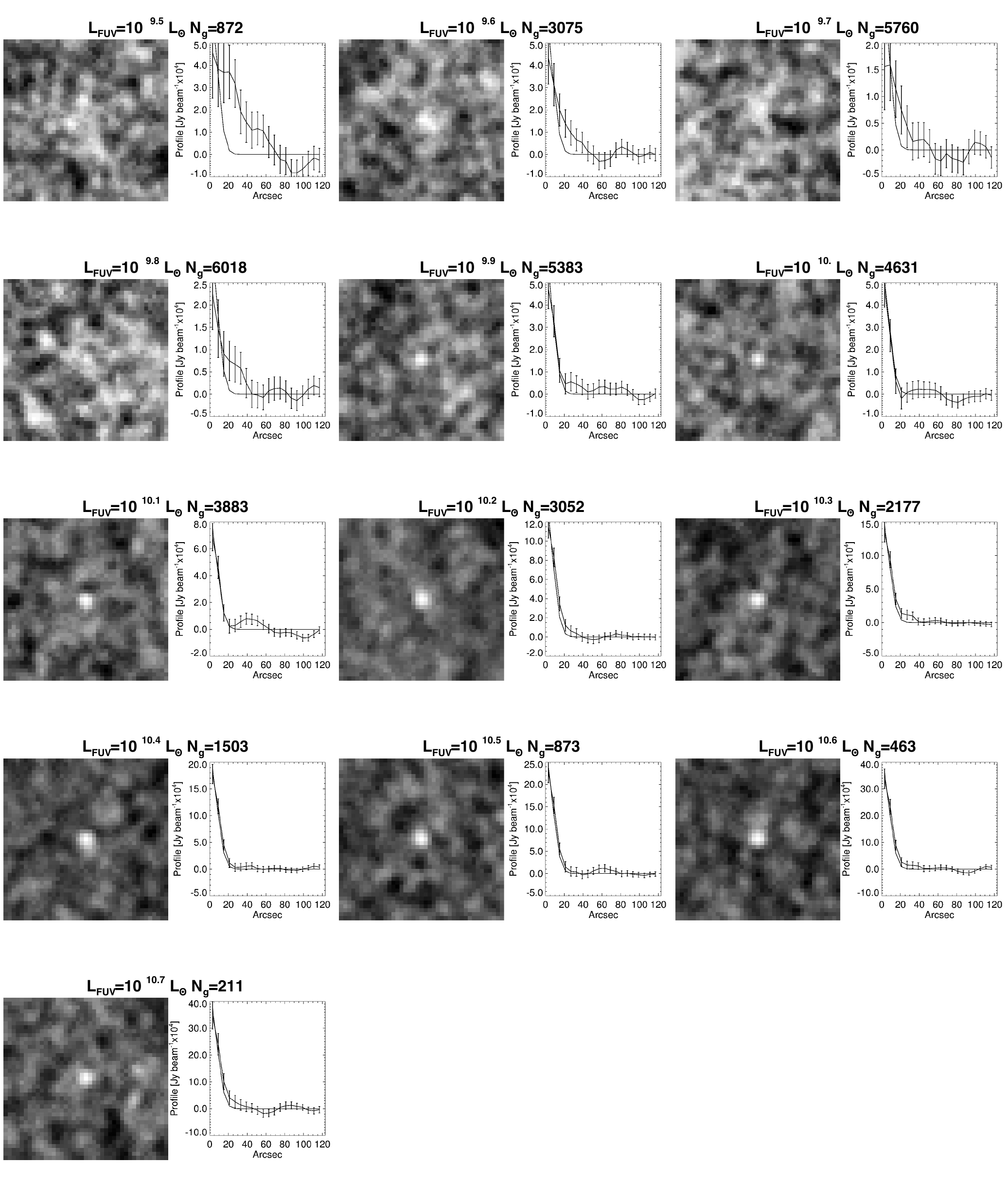}
\caption{Stacking postage stamp images at 250\,$\mu$m in bins of
  $L_{\rm FUV}$. $N_g$ is the number of UV-selected galaxies in each
  bin.}
\label{fig_stamps_Lfuv_250}
\end{figure*}

\begin{figure*}
\includegraphics[width=\hsize]{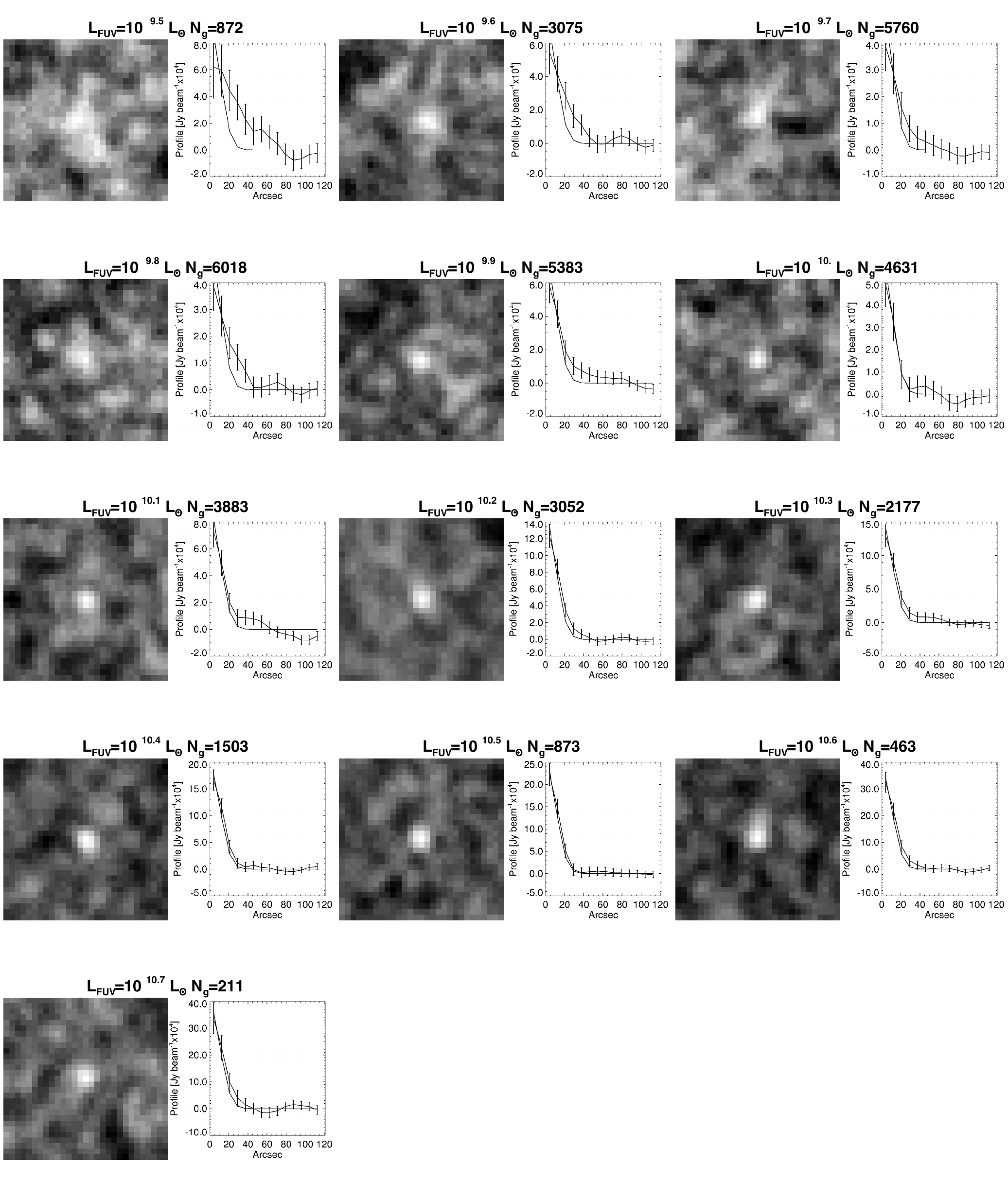}
\caption{Stacking postage stamp images at 350\,$\mu$m in bins of of
  $L_{\rm FUV}$.}
\label{fig_stamps_Lfuv_350}
\end{figure*}

\begin{figure*}
\includegraphics[width=\hsize]{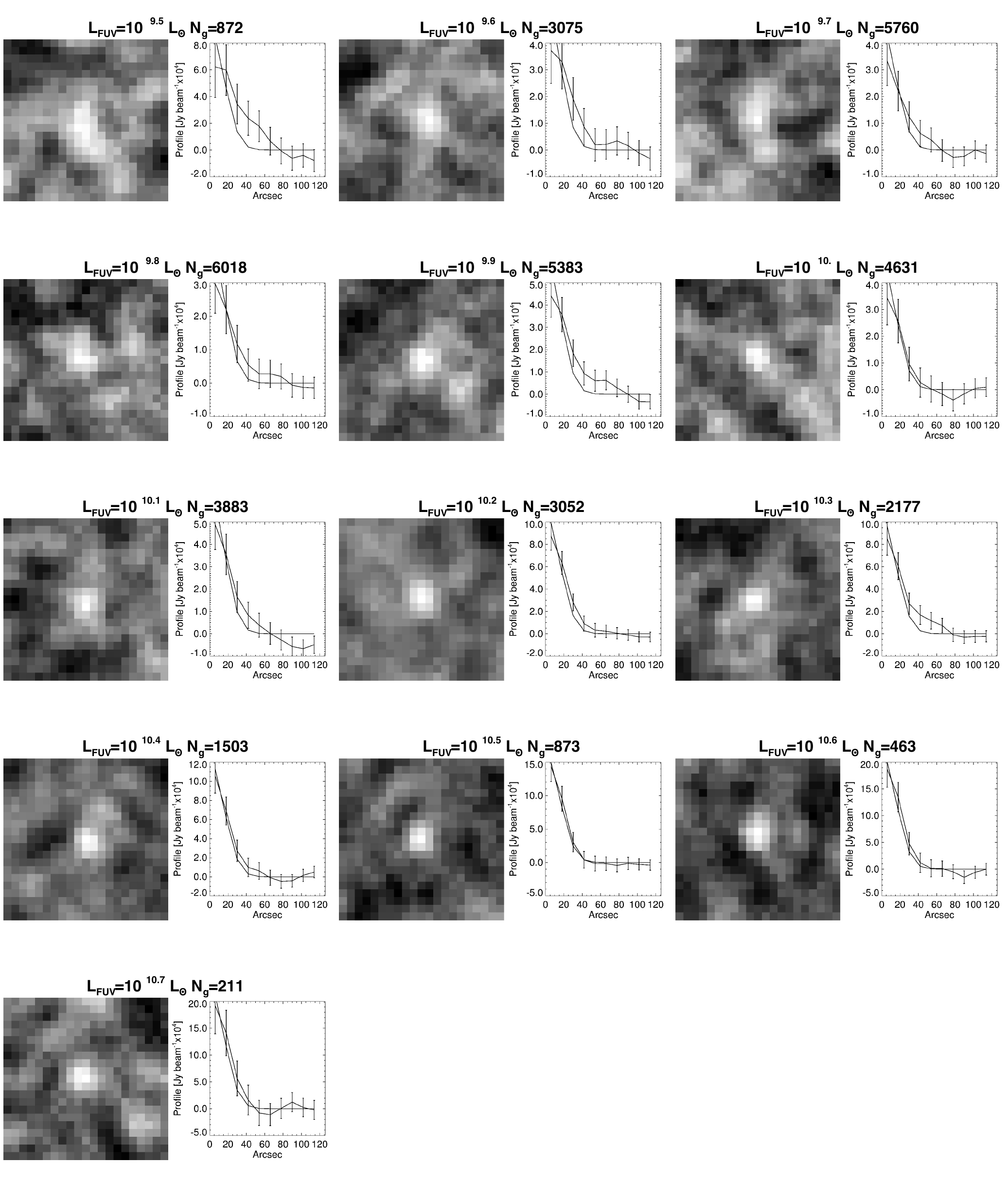}
\caption{Stacking postage stamp images at 500 \,$\mu$m in bins of of
  $L_{\rm FUV}$.}
\label{fig_stamps_Lfuv_500}
\end{figure*}

\begin{figure*}
\includegraphics[width=\hsize]{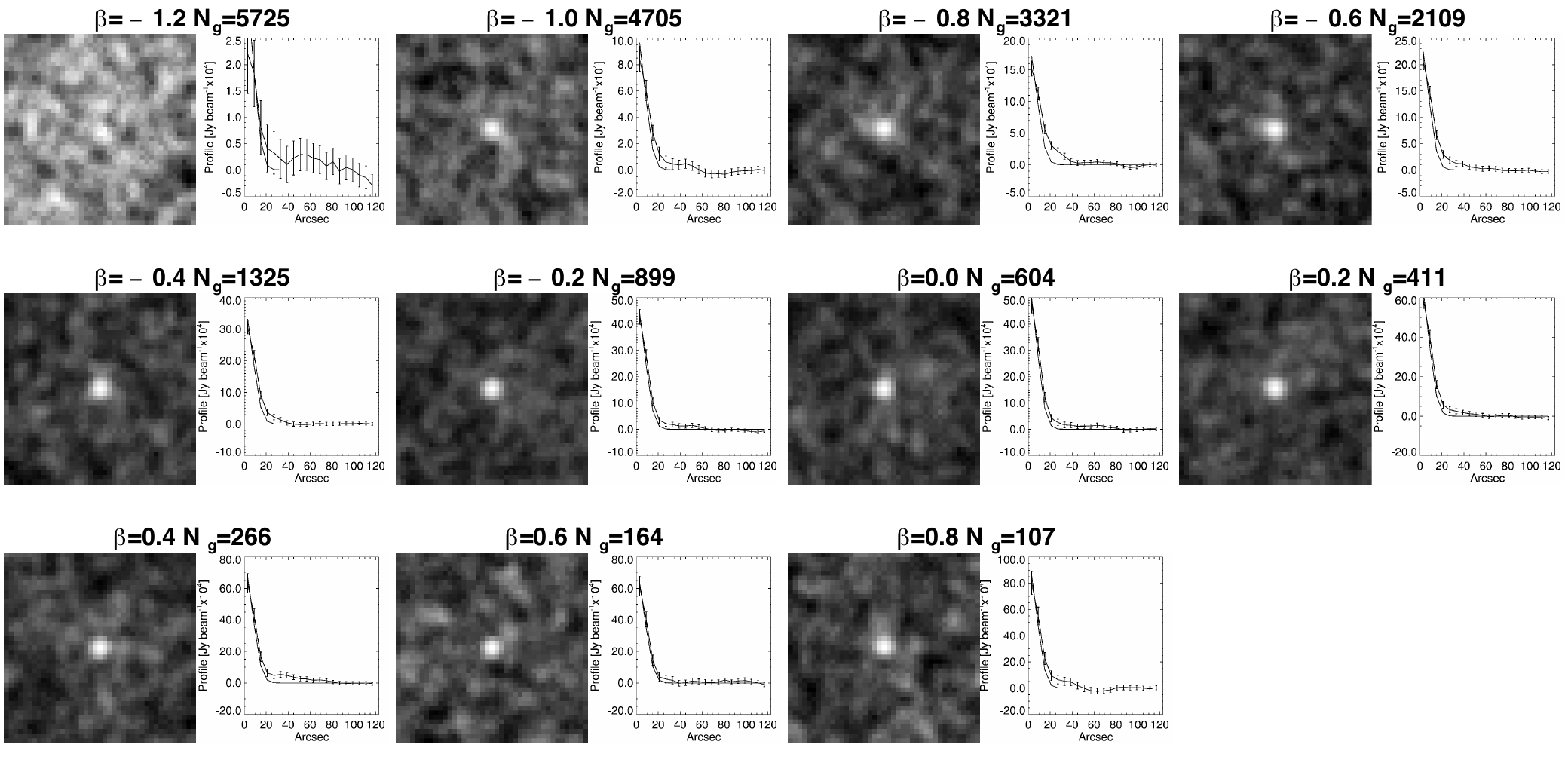}
\caption{Stacking postage stamp images at 250 $\,\mu$m in bins of
  $\beta$, the UV slope.}
\label{fig_stamps_beta_250}
\end{figure*}

\begin{figure*}
\includegraphics[width=\hsize]{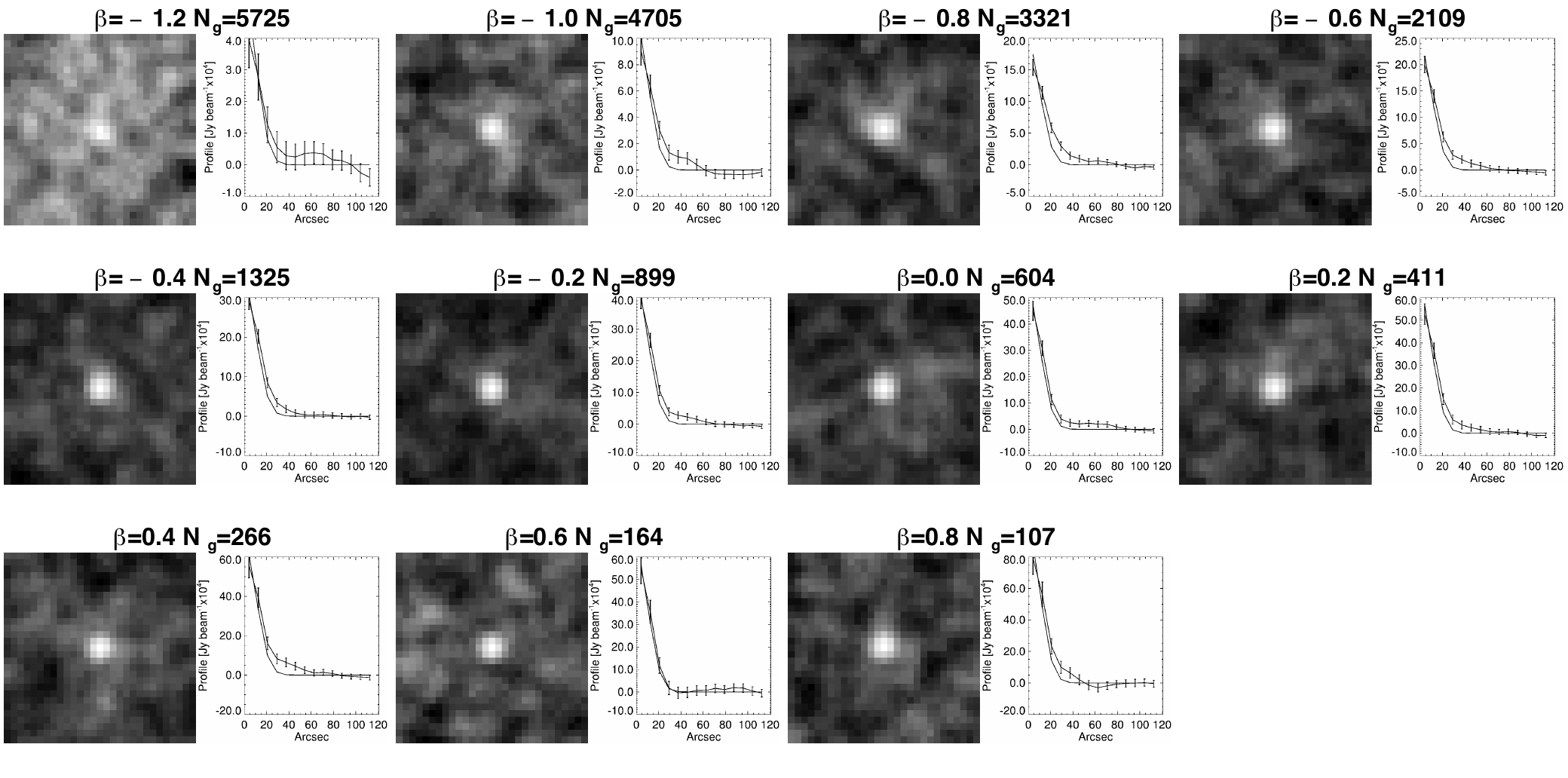}
\caption{Stacking postage stamp images at 350\,$\mu$m in bins of
  $\beta$, the UV slope.}
\label{fig_stamps_beta_350}
\end{figure*}

\begin{figure*}
\includegraphics[width=\hsize]{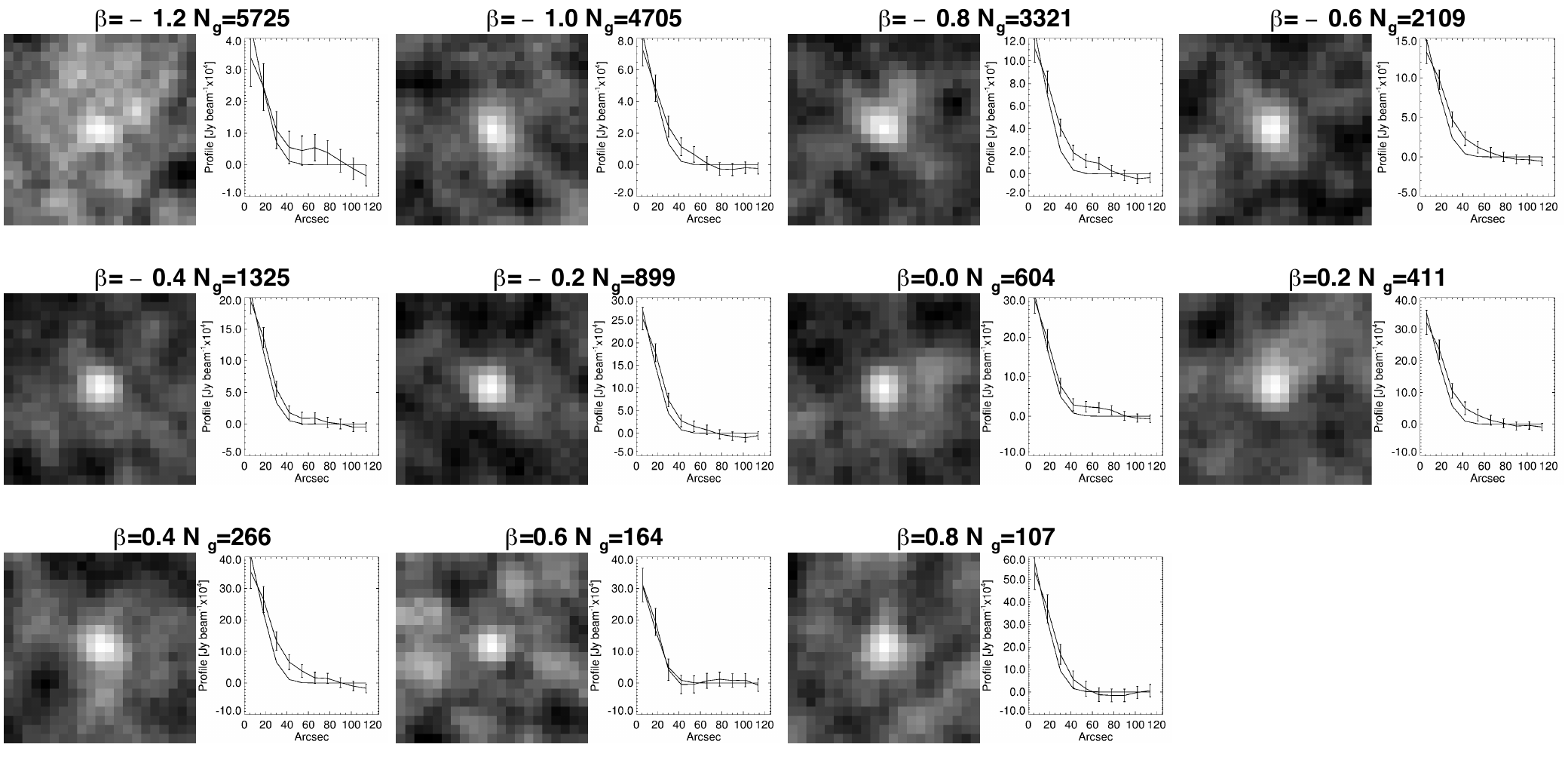}
\caption{Stacking postage stamp images at 500\,$\mu$m in bins of
  $\beta$, the UV slope.}
\label{fig_stamps_beta_500}
\end{figure*}

\section{Comparison with other stacking methods}
\subsection{Comparison of mean stacking versus median stacking}\label{appendix_median_stacking}
\begin{figure}
\includegraphics[width=\hsize]{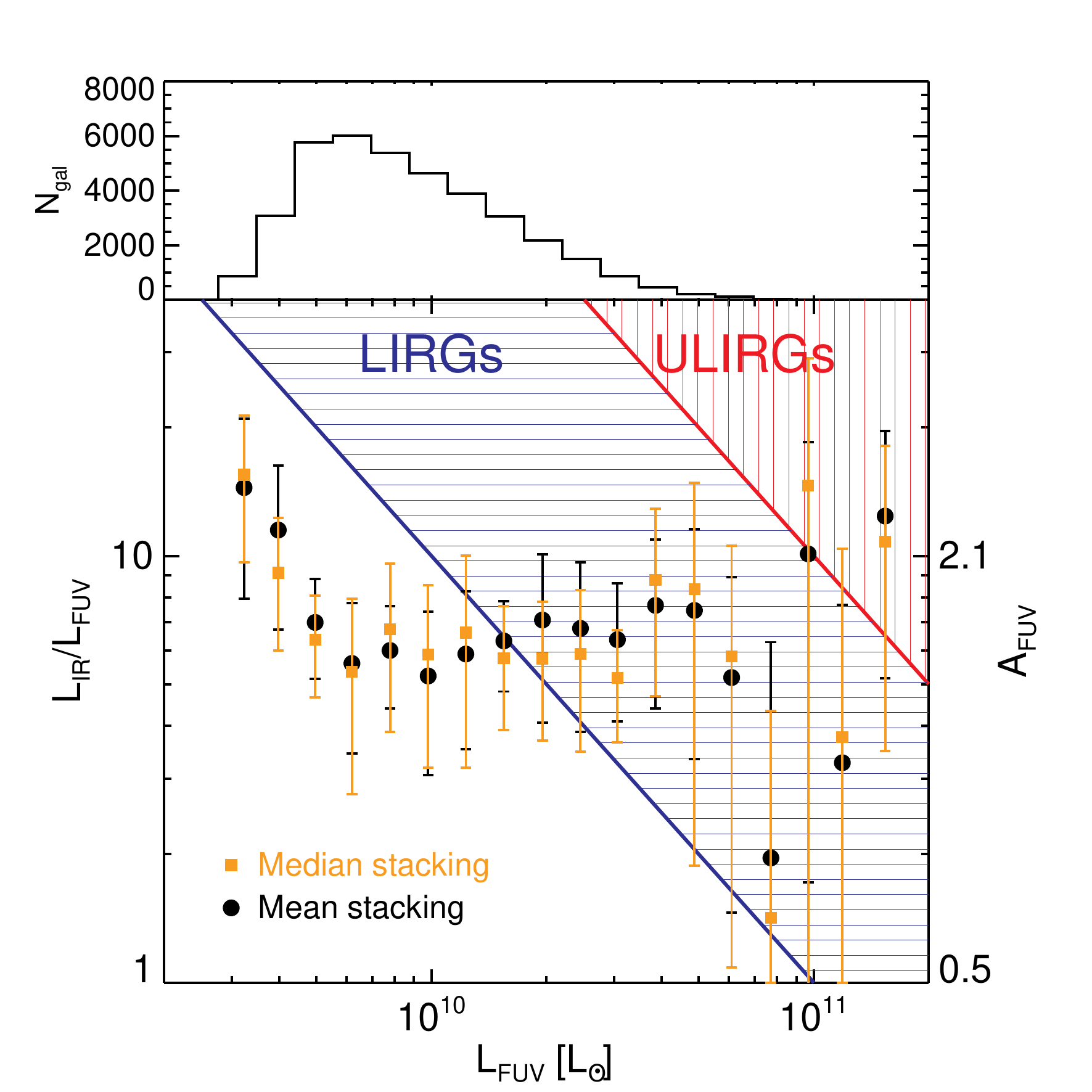} \caption{IR-to-UV
  luminosity ratio as a function of UV luminosity. Circles show
  estimates using the method we describe in this paper. Squares show
  estimates using median stacking, correcting for clustering only. The
  horizontally hatched region represents the locus of LIRGs, while the
  vertically hatched region is for ULIRGs.  The right axis shows the
  equivalent attenuation in the FUV band (in magnitudes), using
  eq. \ref{eq_afuv}. The top panel shows the histogram of galaxies as
  a function of $L_{\rm FUV}$.}
\label{fig_median_stacking}
\end{figure}

In Fig. \ref{fig_median_stacking} we show a comparison of the results
obtained with our method with those obtained using median
stacking. This figure is similar to Fig. \ref{fig_lir_luv_luv}. We
show in black circles the results obtained with our method, and in
orange squares the results obtained with median stacking. For the
latter, we perform median stacking, correcting only for clustering as
described in Sect. \ref{sec_clustering}, but not correcting for
stacking bias (Sect. \ref{incomp}). The agreement between the results
from mean and median stacking is excellent.

\subsection{Comparison of stacking with and without subtraction of detected sources }\label{appendix_cleaning_stacking}
\begin{figure}
\includegraphics[width=\hsize]{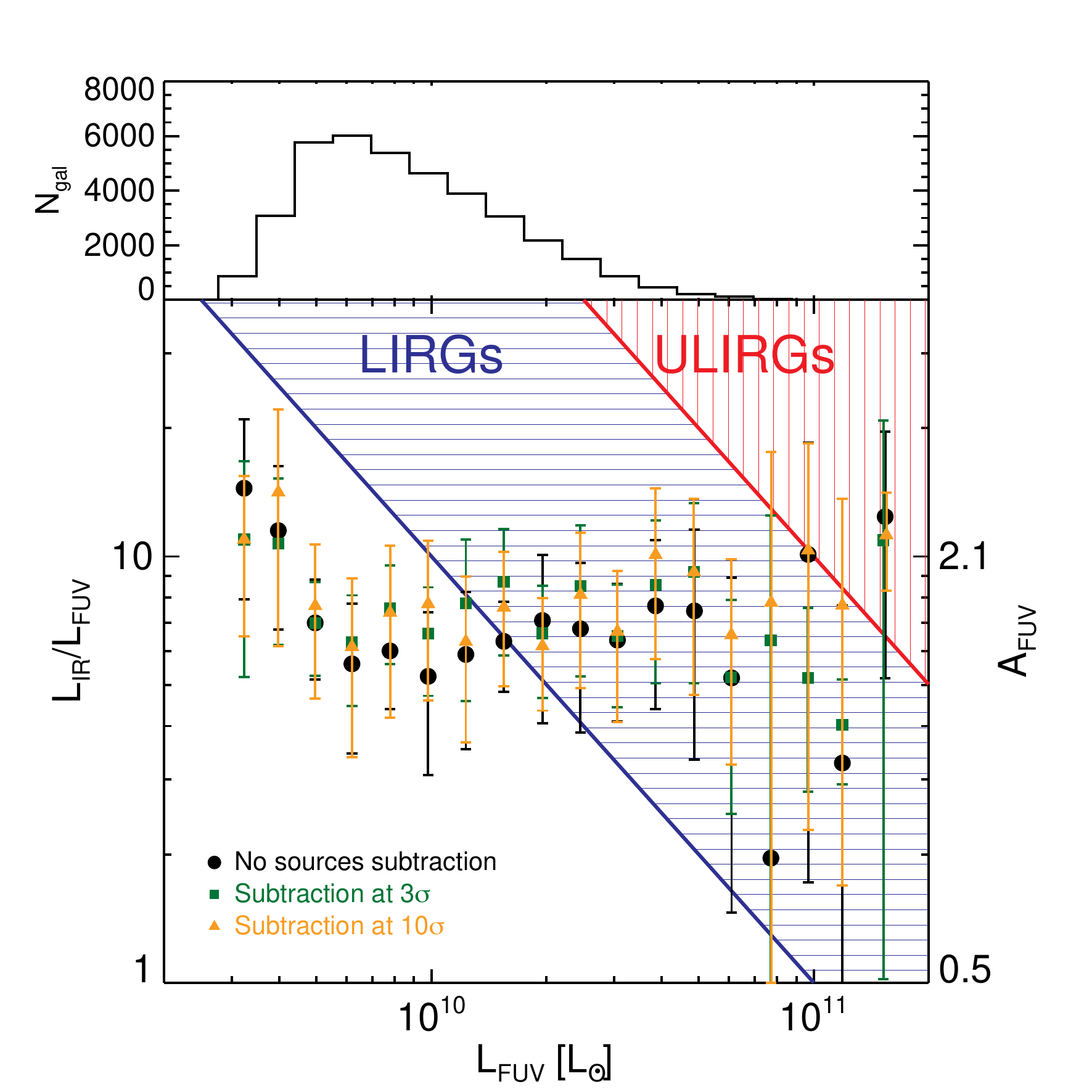}
\caption{IR-to-UV luminosity ratio as a function of UV
  luminosity. Squares (respectively triangles) show the mean stacking
  results obtained while subtracting from the images the sources
  detected at 3$\sigma$ (respectively 10$\sigma$) in each band,
  correcting from clustering only. Circles show the results obtained
  with the method described in this paper. The horizontally hatched
  region represents the locus of LIRGs, while the vertically hatched
  region is for ULIRGs. The right axis shows the equivalent
  attenuation in the FUV band (in magnitudes), using
  eq. \ref{eq_afuv}. The top panel shows the histogram of galaxies as
  a function of $L_{\rm FUV}$.}
\label{fig_cleaning_stacking}
\end{figure}

\begin{figure}
\includegraphics[width=\hsize]{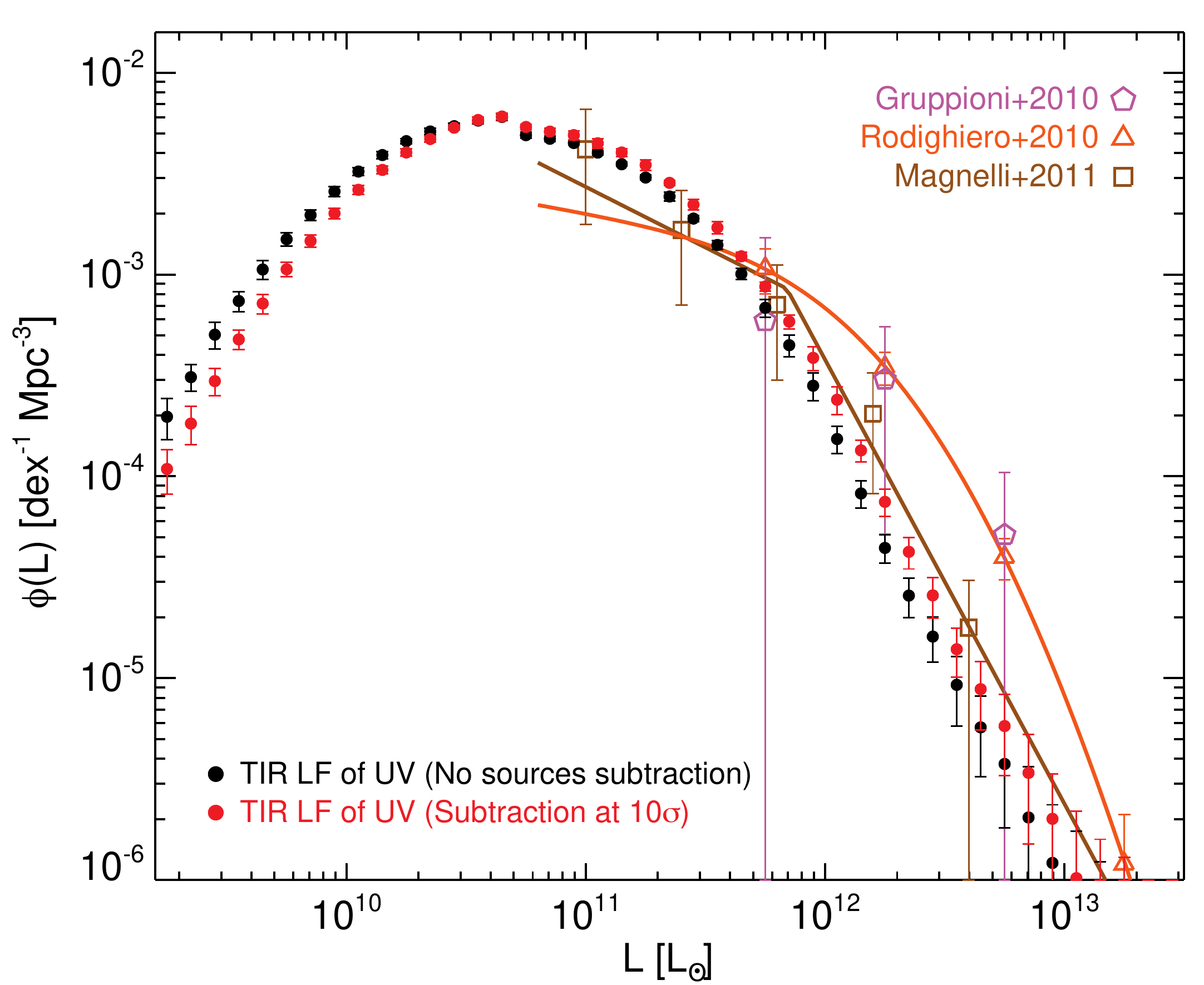}
\caption{IR luminosity functions. The filled circles
    show the IR luminosity function of our sample derived from the
    stacking measurement of the IR to UV luminosity ratio: the black
    circles using the dispersion which reproduces the observed
    $\log(L_{\rm{IR}}/L_{\rm FUV})$ values for detected objects
    (`$\sigma_d$') and our baseline stacking measurements; the red
    circles using the same scenario for the dispersion $\sigma_d$, but
    using the stacking measurements obtained while subtracting from
    the images the sources detected at 10$\sigma$ in each band. Open
    squares show the IR luminosity function of an IR selected sample
    at $1.3<z<1.8$ from \citet{Magnelli_2011}; open triangles the IR
    luminosity function from \citet{Rodighiero_2010} at $1.2<z<1.7$;
    and open hexagons the IR luminosity function from
    \textit{Herschel}/PACS data at $1.2<z<1.7$
    \citep{Gruppioni_2010}.}
\label{fig_lf_cleaning_stacking}
\end{figure}

We compare in Fig. \ref{fig_cleaning_stacking} the results of our
method with another approach where we subtract detected sources from
the images prior stacking. In practice we perform stacking after
subtracting from the images the sources detected either at 3$\sigma$
or 10$\sigma$ in each SPIRE band. We use mean stacking, correcting
only for clustering as described in Sect. \ref{sec_clustering}, but
not correcting for stacking bias (Sect. \ref{incomp}). To obtain the
final estimate, we add to the stacked flux measure the flux of the
UV-selected sources detected with SPIRE which were subtracted. The
resulting fluxes are on average 20 per cent higher
  than those obtained with our main method, even if they are in
  agreement at the $1\sigma$ level, and do not depend significantly
on the threshold used for subtracting sources. We
  show in Fig. \ref{fig_lf_cleaning_stacking} the IR LF built using
  the method $\sigma_d$ (see Sect. \ref{sec_IR_lf}) and these stacking
  measurements, and compare it to the results obtained with our
  baseline stacking results. This IR LF has a slightly larger
  amplitude at the bright end than the one obtained from our baseline
  results, and is hence in better agreement with the LF of
  \citet{Magnelli_2011}.

\end{document}